\documentclass{article}

\input epsf

\begin{document}
\setlength{\unitlength}{1mm}

\begin{titlepage}

\begin{flushright}
LAPTH 759-99\\
November 1999
\end{flushright}
\vspace{1.cm}

\begin{center}
\large\bf
{\LARGE\bf Reduction formalism for dimensionally 
regulated one-loop N-point integrals}\\[2cm]
\rm
{T.~Binoth, J.~Ph.~Guillet and G.~Heinrich$^{a}$ }\\[.5cm]

{\em Laboratoire d'Annecy-Le-Vieux de Physique 
 Th\'eorique\footnote{URM 5108 du CNRS, associ\'ee \`a 
              l'Universit\'e de Savoie.} LAPP,}\\
      {\em Chemin de Bellevue, B.P. 110, F-74941, 
           Annecy-le-Vieux, France}
	
	\medskip   
	    
{\em $^{a}$Department of Mathematics, University of Guelph,\\ 
     Guelph, Ontario N1G 2W1, Canada} \\[3.cm]
      
\end{center}
\normalsize


\begin{abstract}
We consider one-loop scalar and tensor integrals with an arbitrary number
of external legs relevant for multi-parton processes
in massless theories. We present a procedure to
reduce $N$-point scalar functions with generic 4-dimensional
external momenta to box integrals in $(4-2\epsilon)$ dimensions. 
We derive a formula valid for arbitrary $N$
and give an explicit expression for $N=6$.

Further a  tensor reduction method for $N$-point tensor integrals 
is presented. We prove that generically
higher dimensional 
integrals contribute only to order $\epsilon$ for $N\ge 5$.
The tensor reduction can be solved iteratively such that 
any tensor integral is expressible in terms of scalar integrals. 
Explicit formulas are given up to $N=6$.
\end{abstract}

\vspace{3cm}

\end{titlepage}

\section{Introduction}

At future colliders multi-particle/jet final states will become
more and more dominant. A precise theoretical description 
of the QCD reactions is desirable in order to 
have a better control on the backgrounds for
various search experiments. Especially at hadron colliders 
 lots of multi-jet data will be collected which have to be
confronted with theoretical predictions. This motivates 
the accurate calculation of multi-parton reactions.

For tree level calculations the construction of matrix elements
with large numbers of particles in the final state
is well established \cite{treelevelamplitudes}. 
However, tree level results are very unstable with respect to variations of 
the renormalization and factorization scales.
As $2\rightarrow N$ parton cross sections behave as $\alpha_S^N$,
 at least next-to-leading order precision
is necessary to stabilize the predictions.
The knowledge of the respective 
one-loop corrections is thus mandatory.
For $2\rightarrow 3$ parton scattering NLO
contributions have been calculated in 
\cite{EllisSexton,BernDixonKosower,KunsztTrocsanyi},
for $e^+e^-\rightarrow 4\, jets$ see 
\cite{WeinzierlKosower,CampbellCullenGlover}.

The main technical difficulties for constructing amplitudes 
consist in the treatment of the occurring $N$-point scalar and tensor 
integrals. Because of infrared (IR) divergencies the standard methods of
\cite{PassarinoVeltman,Oldenbourg,VanNeervenVermaseren} 
are not directly applicable. The authors of
\cite{BernDixonKosower}  used a formalism working in
Feynman parameter space. The generation of parameter integrals with 
nontrivial numerators connected to tensor integrals was
done by differentiation techniques. The formalism produced formally 
higher dimensional integrals which in the end canceled out. This
cancellation had to be shown by explicit calculation. In this 
paper we present a proof that this cancellation mechanism
is true for arbitrary $N$. Furthermore, 
a scalar reduction formula was derived in \cite{BernDixonKosower}
which relates scalar
integrals in different space-time dimensions to each other.
For 4-dimensional external momenta 
the formula could not be shown to hold true in general. 
The problem stemmed from the presence of singular matrices.
We rederive the formula and treat the necessary inversions
of these matrices
by using so-called pseudo--inverse matrices.  
To keep the external momenta in 4 dimensions
is advantageous because it allows to use helicity techniques \cite{Signer}.

Another formalism based on the generation of tensor integrals 
by differentiation of  
scalar integrals serving as generating functionals has been proposed in 
\cite{Davydychev,FleischerJegerlehnerTarasov}. The reduction of the scalar
integrals works basically in the same way as in  \cite{BernDixonKosower}
with the same problems in the case of 4-dimensional kinematics.
Their method has been designed for massive integrals and the applicability 
to the massless case has not been worked out. The authors state
that a regulator mass is needed for infrared divergencies in the
case of 4-dimensional kinematics. 

A generalization of the Passarino-Veltman techniques
dealing with the problem of vanishing Gram determinants
has been developed in \cite{DevarajStuart}. The authors 
show explicitly how to reduce box tensor integrals to scalar
integrals.
Another approach which uses helicity 
methods for the reduction of tensor integrals
to scalar integrals has been discussed in \cite{Pittau,Weinzierl}.

The complete reduction to scalar integrals 
is also  possible in our tensor reduction scheme.
Our formalism is in a sense the generalization
of the Passarino-Veltman methods used for the calculation of electroweak
radiative corrections \cite{Denner} to the massless case.

The paper is organized as follows. In section 2 we rederive a reduction
formula for scalar integrals. We concentrate on the case of 
4-dimensional kinematics from the start and show that any
$N$--point scalar integral with $N\ge 6$ is a linear combination of pentagon
integrals which in turn are  combinations of box integrals plus 
terms of ${\cal O}(\epsilon)$. We give an explicit formula for
the 6-point function with all external legs on-shell.
In section 3 we derive our tensor reduction formalism which 
combines Passarino-Veltman--like methods with Feynman parameter
space techniques. We prove that for generic 4-dimensional 
kinematics all higher dimensional 
$N$-point functions drop out for arbitrary $N\ge 5$ and  generalize 
the reduction methods of  \cite{BernDixonKosower,CampbellGloverMiller}
to arbitrary $N$.
We construct a hierarchy of tensor formulas up to $N=6$ and rank $\le N$
which can be solved by iteration. The explicit expressions
and the derivation of some basic formulas
are given in the appendix.

\section{Reduction formula for massless scalar integrals}
In this section we will first derive a scalar reduction formula
valid for an arbitrary number of external legs. Then we will
use these formulas to derive explicit representations
for scalar integrals up to $N=6$.
\subsection{Derivation}
Consider a one-loop, scalar $N$-point function with massless
propagators for $N\ge 4$. 
If all legs are off-shell the integral is finite and can
be treated in four dimensions \cite{VanNeervenVermaseren}.
If at least one external leg is massless it is infrared divergent
and needs a regulator. 
In the framework of dimensional regularization, 4-dimensional methods
are not applicable anymore. We work in  
$n=4-2\epsilon$ dimensions in the following
with the external momenta kept in four dimensions.

\begin{figure}[h]
\hspace{4.5cm}
    \epsfxsize = 6cm
    \epsffile{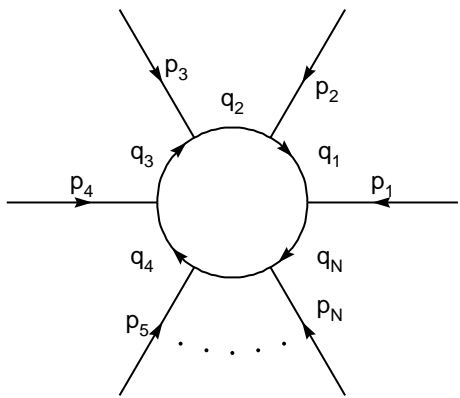}
\caption{\label{npo-graph}{\em $N$-point graph.}}
\end{figure}

With the 
momentum flows as indicated in 
Fig.~\ref{npo-graph}, we define the propagator momenta as $q_l = k - r_l$
with $r_l=p_l+r_{l-1}$ for $l$ from $1$ to $N$ and  $r_0=r_N$. 
Momentum conservation allows to choose $r_N=0$.
 
The corresponding analytic expression in momentum and
Feynman parameter space is
\begin{equation}
I_N^n(R) = \int d\kappa \; \frac{1}{\prod^N_{l=1} q_l^2} = 
(-1)^N \Gamma(N-n/2) \int_{0}^{\infty} d^Nz 
\frac{\delta(1-\sum_{l=1}^N z_l)}{(z\cdot S \cdot z)^{N-n/2}} 
\label{EQdefi}
\end{equation}
Herein $R=(r_1,\dots,r_{N})$, $d\kappa = d^n k/(i\pi^{n/2})$.
The kinematic information is contained in the matrix $S$
which is related to the Gram matrix $G$ by
\begin{eqnarray} \label{EQStoG}
S_{kl}&=&-\frac{1}{2}(r_l-r_k)^2 = \frac{1}{2}( G_{kl} - v_l -v_k )\\
G_{kl}&=&2\,r_k\cdot r_l\; ,\quad v_k=G_{kk}/2 \; ,
\quad k,l=1,\ldots,N \nonumber
\end{eqnarray}
Although it is well known 
\cite{BernDixonKosower,FleischerJegerlehnerTarasov,CampbellGloverMiller} 
that the $N$-point integral
can be split into a finite, $(6-2\epsilon)$-dimensional integral and
a part with less external legs containing the infrared poles,
we want to present our derivation which later allows to deal with
the problem of vanishing Gram determinants in a transparent manner.
As an ansatz we write (\ref{EQdefi}) as a sum of (one-propagator)
reduced diagrams and a remainder.
\begin{eqnarray}
I_N^n & = & I_{div} + I_{fin} = \int d\kappa \; 
\frac{\sum^N_{l=1} b_l \, q_l^2}{\prod^N_{l=1} q_l^2} +
 \int d\kappa \; 
\frac{\left[ 1 - \sum^N_{l=1}  b_l \, q_l^2 \right] }{\prod^N_{l=1} q_l^2}
\label{EQdivetfin}
\end{eqnarray}
We want to show in the following that one can find coefficients
$b_l$ such that $I_{fin}$ contains no IR poles.
In standard Feynman parametrization one gets with 
$D = \sum^N_{l=1} z_l \, q_l^2$
\begin{eqnarray}
I_{fin} & = & \Gamma(N) \, \int^\infty_0 d^Nz \;  
\delta(1- \sum^N_{i=1} z_i) \; \int d\kappa  \; 
\frac{\left[ 1 - \sum^N_{l=1}  b_l \, q_l^2 \right] }{D^N} 
\label{EQfinpart1}
\end{eqnarray}
and after a shift $k =\tilde k + \sum_{l=1}^N z_l \, r_l$
\begin{eqnarray} 
I_{fin} & = &\Gamma(N) \, \int^\infty_0 d^Nz \; 
\delta(1- \sum^N_{l=1} z_l) \; \int d\tilde\kappa \; 
\frac{\left[ 1 - \sum^N_{l=1} b_l \,\tilde{q}_l^2 \right]}{(\tilde{k}^2-M^2)^N}
\label{EQfinpart2}
\end{eqnarray}
with
\begin{eqnarray}
M^2 = z\cdot S\cdot z \, ,\quad 
\tilde{q}_j & = & \tilde{k} - \sum_{l=1}^N (\delta_{lj} - z_l) \; r_l
\label{EQdiffeq}
\end{eqnarray}
Now the term in square brackets in Eq.~(\ref{EQfinpart2}) can be written as
\begin{eqnarray}
\left[1 - \sum^N_{i=1} b_i \, \tilde{q}_i^2 \right]
& = & - (\tilde{k}^2+M^2 ) \sum\limits_{j=1}^{N} b_j + 
\sum\limits_{j=1}^{N} z_j \Bigl(  1+2 ( S\cdot b)_j  \Bigr)
\label{EQdefbracket}
\end{eqnarray}
If now the equations
\begin{eqnarray}
(S\cdot b)_j = -\frac{1}{2} \,, \quad j=1,\dots,N
\label{EQcoeffbi}
\end{eqnarray}
are fulfilled,
the second term on the right-hand-side of (\ref{EQdefbracket}) vanishes
and one finds 
\begin{eqnarray}
I_{fin} & = & - \Gamma(N) \, \left( \sum_{l=1}^N b_l \right) \;
 \int^\infty_0 d^Nz \; \delta(1- \sum^N_{l=1} z_l) 
\int d\tilde\kappa \; 
\frac{\tilde{k}^2+M^2}{(\tilde{k}^2-M^2)^N} 
\label{EQifin1}
\end{eqnarray}
Finally the loop momentum integration gives  
\begin{eqnarray}
I_{fin} & = &  \left( \sum_{l=1}^N b_l \right) 
(-1)^{N+1} \, \Gamma(N-1-\frac{n}{2}) 
\,(N-n-1) \,\int^\infty_0  d^Nz \; 
\frac{\delta(1- \sum^N_{l=1} z_l)}{(M^2)^{N-(n+2)/2}} \nonumber \\
&=& -\left( \sum_{l=1}^N b_l \right) (N-n-1)\, I_N^{n+2}
\label{EQifin2}
\end{eqnarray}
The $(6-2\epsilon)$-dimensional integral is IR finite, as can be  seen
by a power-counting argument 
in the corresponding momentum integral.

It remains to solve Eq.~(\ref{EQcoeffbi}).
In the case of non-exceptional 4-dimensional kinematics, 
 $rank(S)=min(N,6)$ holds.
For $N\le 6$ one has $b_l=-1/2\sum_{k=1}^{N} S^{-1}_{kl}$.
In the case $N=5$, $I_{fin}$ contains a factor $(N-1-n)$ 
which is ${\cal O}(\epsilon)$. As is well known, pentagon integrals
are just a sum of box integrals up to a
remainder which drops out in phenomenological applications. 
In the case $N> 6$ and 4-dimensional kinematics, 
Eq.~(\ref{EQcoeffbi}) does not have a unique solution.
To clarify this point it is useful
to rewrite Eq.~(\ref{EQcoeffbi}). Using momentum conservation, 
$r_N=0$, Eq.~(\ref{EQcoeffbi})
leads to the following equations
\begin{eqnarray}\label{EQlinearequation}
\sum\limits_{l=1}^{N-1} G_{kl} b_l = v_k B_N \quad, \quad 
\sum\limits_{l=1}^{N-1} v_{l} b_l = 1 \quad, \quad 
B_N = \sum\limits_{l=1}^{N}  b_l
\label{EQbcoeffsG}
\end{eqnarray}
Herein $G$ is now the Gram-matrix of the vectors $r_{l=1\dots N-1}$.
In the case of four dimensional kinematics it 
is at most of rank 4 for all $N\ge 5$. 
Generically any four vectors out of ${r_1,\dots,r_{N-1}}$ are 
linearly independent. 
Such a configuration is called non-exceptional in the following.
Choosing four linearly independent vectors
$E^{\mu}_{l=1,\dots,4}$ as a basis of the physical Minkowski space,
one can define a coefficient matrix $R$ and a Gram matrix $\tilde G$
made out of these basis vectors 
\begin{eqnarray}
r_j^{\mu} = \sum\limits_{l=1}^{4} R_{lj} E^{\mu}_l \, , 
\quad\tilde G_{jk} = 2 \,E_j\cdot E_k
\label{EQbasis}
\end{eqnarray}
The Gram matrix $G$ is expressible as $G=R^T \,\tilde G\, R$. 
  
Based on (\ref{EQbasis})  one can construct the most
general solution for the case $\det (G)=0$ by means of pseudo-inverse matrices.
This concept will also be useful for the tensor integrals.
 
A pseudo-inverse $H$ to $G$ is defined by the conditions 
$H\, G\, H=H$ and $G\, H\, G=G$. 
Given a pseudo-inverse matrix $H$ to $G$, the following statement 
can be proven. A solution to the linear equation 
$G\cdot x=y$ exists if and only if 
$y=G\, H\cdot y$.
Then the most general solution can be written as
$x=H\cdot y+(1_{N-1}-H\, G)\cdot u$, where the last term --- with  
$u\in R^{N-1}$ arbitrary ---
is parametrizing the solutions of the homogeneous equation.
As $G$ is  symmetric, its pseudo-inverse is uniquely defined \cite{Koecher}. 
The pseudo-inverse to $G$ is given by
\begin{equation}
H = R^T \,(R\,R^T)^{-1} \,\tilde G^{-1}\, (R\,R^T)^{-1}\, R
\label{EQdefpseudoinvers}
\end{equation}   
where $R$ is defined in (\ref{EQbasis}).
For $N=5$ and non-exceptional external momenta 
the vectors $r_{1},\ldots,r_{4}$ are the natural basis. 
Then $R=1_{4}$, $\tilde G=G$, $H=G^{-1}$.
Note that the concept of the pseudo-inverse always allows the inversion of 
linear equations containing the Gram matrix, even if the external momenta are
exceptional\footnote{Clearly one could also solve Eq.~(\ref{EQcoeffbi})
by means of the pseudo-inverse to $S$.
With $O_S$ defining the orthogonal transformation 
which diagonalizes $S$  
and the non-zero eigenvalues $\lambda_1,\dots,\lambda_6$ of $S$, 
the pseudo-inverse to $S$
 is given by $O^T_S\cdot D^{inv}_S \cdot O_S$, where 
$D^{inv}_S=diag(1/\lambda_1,\dots,1/\lambda_6,0,\dots,0)$. 
Its computation is algebraically more involved than the
computation of $H$ though.}.

If the Gram matrix is not of maximal rank, 
the condition $v=G\, H\cdot v$ for the existence of a
solution of the inhomogeneous 
equation in (\ref{EQbcoeffsG}) is never fulfilled. 
This can be seen by diagonalizing
the symmetric matrix $G\, H$ by an orthogonal transformation $O$,
$G\, H= O^T \, D\, O$, where $D$ is a diagonal matrix with
elements $(1,1,1,1,0,\dots ,0)$. 
Now it is evident that in general
$O\cdot v \not = D\, O\cdot v$.
Thus a solution to Eq.~(\ref{EQbcoeffsG}) exists only for the case $B_N=0$.
The solution of the homogeneous equation spans a $(N-5)$-dimensional
space which is just the kernel of the Gram matrix. 
It can be parametrized by $(N-5)$  vectors
$U^{(1,\dots,N-5)}$. 
Defining 
\begin{eqnarray}
K=1_{N-1}-H\, G= 1_{N-1} - R^T\, (R\, R^T)^{-1}\, R \;, 
\end{eqnarray}
one has 
$K\cdot v \in ker(G)$. Now one can choose
$U^{(N-5)}=K\cdot v/(v\cdot K\cdot v)$ parallel to $v$ and
the others orthogonal, $v\cdot U^{(k=1,\dots,N-6)}=0$.
A general vector in $ker(G)$ is then parametrized by
$U=\sum_{k=1}^{N-6} \beta_k  U^{(k)} +\alpha U^{(N-5)}$ and 
the general solution is  given by
\begin{eqnarray}\label{EQparametrizedsolution}
b_i  &=& \frac{(K\cdot v)_i +
\sum_{k=1}^{N-6} \beta_k  U^{(k)}_i }{v\cdot K \cdot v}  
\quad ,\quad i=1,\dots,N-1 \nonumber\\
B_N &=& \sum_{k=1}^{N} b_k = 0
\end{eqnarray}
where $\alpha =1$ is imposed by $v\cdot b=1$.
Thus for $N=6$ the solution to (\ref{EQbcoeffsG}) is unique
and therefore equal to the one defined by the inverse of the matrix $S$.
Eq.~(\ref{EQparametrizedsolution}) 
proves constructively that for all $N\ge 6$, 
$I_N^n$ can be expressed in terms of $(N-1)$-point functions without
the higher dimensional remainder term. 
This is a consequence of the linear dependence of
propagators if the external momenta are 4-dimensional.
For the special case $N=7$ this has already been demonstrated in 
\cite{BernDixonKosower}.
Obviously one can
choose the $\beta$'s to eliminate $(N-6)$ $b$'s from the set 
$\{b_1,\dots , b_{N-1}\}$.
Doing so one observes that
$I_N^n$ can be expressed by only 6 $(N-1)$-point graphs for arbitrary $N\ge 6$.
Here we make contact to
a result in \cite{VanNeervenVermaseren,Melrose}, where a similar  relation
has been derived for the IR finite case in integer dimensions by 
using 4-dimensional Schouten identities. 

For $N\le 6$ one has $\det(S)\not =0$ and the following relation 
(see also \cite{BernDixonKosower}) holds
\begin{eqnarray}\label{EQdetGformula}
\sum\limits_{l=1}^N b_{l} =-\frac{1}{2}\sum\limits_{l,k=1}^N S^{-1}_{lk} = -\frac{\det(G)}{2^{N}\det(S)}
\end{eqnarray}
which shows that the vanishing of the finite remainder terms (\ref{EQifin2})
is related to the vanishing of the Gram determinant.

In summary, with the definition of reduced graphs
\begin{eqnarray}\label{EQreducedgraph}
I^n_{N-p,l_1,\dots,l_p} =\int d\kappa \; \frac{ 
\prod^p_{m=1} q_{l_m}^2 }{\prod^N_{l=1} q_l^2}
\quad ,\quad I^n_{N-p,N-l_1,\dots,N-l_p} =\int d\kappa \; \frac{ 
\prod^p_{m=1} (q_N^2-q_{l_m}^2) }{\prod^N_{l=1} q_l^2}\: 
\end{eqnarray}
the scalar reduction formula for a regular matrix $S$ is
\begin{eqnarray}\label{EQscalarreductionNT}
I_N^n &=& -\frac{1}{2}\sum \limits_{k,l=1}^{N} S^{-1}_{kl} \,
  I_{N-1,l}^n  + (N-n-1) \,\frac{\det(G)}{2^{N}\det(S)}\, I_N^{n+2} 
  \quad , \quad \det(S)\not =0.
\end{eqnarray}
In the case of a singular matrix $S$, as outlined above, we can always
achieve a representation without higher-dimensional integrals.
\begin{eqnarray}\label{EQscalarreductionT}
I_N^n &=& - \frac{1}{v\cdot K\cdot v}\sum \limits_{l=1}^{N-1} 
\Bigl( (K\cdot v)_l+\sum_{k=1}^{N-6} \beta_k  U^{(k)}_l\Bigr) 
I_{N-1,N-l}^n  \quad ,\quad \det(S)=0.
\end{eqnarray}
Remember that the $(N-6)$ $\beta$'s are free parameters and that
the $U^{(k=1,\dots ,N-6)}$ are lying in the kernel of $G$ and are
orthogonal to $v$. The case $N=6$ is special in the sense that
it is of the form  (\ref{EQscalarreductionNT}) 
with vanishing higher-dimensional term.\\
By applying the above formulas iteratively,
any $N$-point function can be reduced
to linear combinations of box integrals 
plus irrelevant ${\cal O}(\epsilon)$
terms in a constructive way.

\subsection{Application}  
We will now discuss the formulas for the cases up to 
$N=6$ in more detail. 
In these cases one has $\det(S)\not =0$ and relation (\ref{EQdetGformula}) 
holds. Note that the vanishing of the Gram determinant 
for $N<6$ typically occurs
at a border of the respective phase space.
In the case $N=3$ $(4,5)$ and $\det(G)=0$ the scalar integrals
are just sums of 2-point, (3-point,4-point) functions respectively,
an observation made some time ago by Stuart \cite{DevarajStuart}
and reflected by Eq.~(\ref{EQscalarreductionNT}).
This fact can be used as a guideline to define nontrivial groupings of
$(N-1)$- and $N$-point functions
which is helpful for numerical purposes \cite{CampbellGloverMiller}. 
The other dangerous determinant is $\det(S)$ which vanishes 
if $N\ge 7$.
We note that in this case
our formulas are mathematically well-defined and that for the 
generation of explicit expressions one just has to apply 
formula (\ref{EQscalarreductionT}). 
Due to the freedom
to choose the parameters $\beta$, the
representation of the $N$-point function
in terms of reduced integrals is not unique as 
it is the case for $N\le 6$. This is a reflection of the fact that
the respective reduced integrals are not linearly independent.

As the reduction formulas were derived under the assumption
$N\ge 4$, we discuss now the special cases $N=1,2,3$ first.
We will represent the kinematic information in terms of
the matrix $S$. The relation to the Gram matrix 
is defined in Eq.~(\ref{EQStoG}).

\subsubsection*{Cases N=1,2:}
As massless tadpole integrals are zero in dimensional regularization, $I_1^n=0$.
The case $N=2$ is trivial in the sense that reduced graphs are tadpoles.
The formula simply gives a 
relation between $n$- and $(n+2)$-dimensional two-point functions. If $p_1$ 
is the external momentum with $p_1^2=m_1^2$,  one obtains
\begin{eqnarray}\label{EQSR2}
I_2^n = 2 \frac{(n-1)}{m_1^2} \, I_2^{n+2}
\end{eqnarray}
For vanishing Gram determinant, i.e. $m_1^2=0$, $I_2^n=0$ in dimensional 
regularization. 
\subsubsection*{Case N=3:}
The matrix $S$ generally looks like ($p_k^2=m_k^2$):
\begin{eqnarray}
S = -\frac{1}{2} \left(  \begin{array}{ccc} 0     & m_2^2 & m_1^2 \\
                                             m_2^2 & 0     & m_3^2 \\
                                             m_1^2 & m_3^2 & 0 
                       \end{array}  \right)
\end{eqnarray}
Note that in the cases of one or two external lines on-shell,
$S$ is not of maximal rank. 
If for example $m_1^2=m_2^2=0$, or $m_1^2=0,m_2^2\not = m_3^2$,  
there exists no solution to (\ref{EQcoeffbi}) 
and no reduction is possible. 
%
%
If all legs are off-shell one finds a solution which relates an 
$(n+2)$-dimensional  off-shell triangle to an $n$-dimensional one.
\begin{eqnarray}\label{EQSR3}
I_3^n = -\frac{1}{2}
\sum\limits_{l,k=1}^3 S^{-1}_{lk} I_{2,l}^n 
+ \frac{\det(G)}{8\det(S)} (2-n) I_3^{n+2}
\end{eqnarray}
We will see later that the $(n+2)$-dimensional off-shell triangle
appears in the tensor reduction  of rank $\ge 3$ tensor 
4-point functions. 

As any $N$-point integral in $(4-2\epsilon)$ dimensions 
can be reduced to triangles plus a finite remainder, 
we see that they are the atoms of any scalar reduction formula. 
Moreover, at the one-loop level they define an IR counterterm
structure to any $N$-point function\footnote{This is actually to be expected
from looking at the reduced diagrams (obtained by shrinking
all finite propagators, if the loop momentum
becomes soft/collinear) of the one-loop $N$-point function
corresponding to solutions of the Landau equations. The respective
reduced  diagrams are just the reduced diagrams of triangle graphs.}.

The explicit expressions for the three types of
triangle graphs may be found for example 
in \cite{CampbellGloverMiller}. 

\subsubsection*{Case N=4:}
With $(p_1+p_2)^2=s$, $(p_2+p_3)^2=t$, one has
\begin{eqnarray}
S = -\frac{1}{2}\,\left(  \begin{array}{cccc} 
0     &  m_2^2 &  t     &  m_1^2 \\
m_2^2 &    0   &  m_3^2 &  s     \\ 
t     &  m_3^2 &  0     &  m_4^2 \\
m_1^2 &    s   &  m_4^2 &  0  
\end{array}  \right)
\end{eqnarray}
$S$ is of maximal rank for all on-shell cases as long as 
$s$ and $t$ are non-zero.
The reduction formula (\ref{EQscalarreductionNT}) applies and
relates a 4-point function in $n$ dimensions with triangles and 
$(n+2)$-dimensional 4-point functions. 
\begin{eqnarray}\label{EQSR4}
I_4^n =  -\frac{1}{2}
\sum\limits_{l,k=1}^4 S^{-1}_{lk}  I_{3,l}^n 
+  \frac{\det(G)}{16\det{(S)}}(3-n) I_4^{n+2}
\end{eqnarray}
For $n=4-2\epsilon$ the triangles carry all the infrared poles
whereas the $(6-2\epsilon)$ dimensional part is a finite remainder.
It can be calculated directly by setting $\epsilon=0$. 
Explicit expressions for the box integrals
can be found in \cite{BernDixonKosower,CampbellGloverMiller}.

\subsubsection*{Case N=5:}
With $(p_j+p_{j+1})^2=s_{j,j+1}$ $(j\mbox{ mod } 5)$ one has
\begin{eqnarray}
S = -\frac{1}{2}\,\left(  \begin{array}{ccccc} 
0      &  m_2^2 &  s_{23} &  s_{51} &  m_1^2   \\
m_2^2  &    0   &  m_3^2  &  s_{34} &  s_{12}  \\ 
s_{23} &  m_3^2 &  0      &  m_4^2  &  s_{45}  \\
s_{51} &  s_{34}&  m_4^2  &  0      &  m_5^2   \\
m_1^2  &  s_{12}&  s_{45} &  m_5^2  &   0   
\end{array}  \right)
\end{eqnarray}
For $N=5$ the higher dimensional integrals come with a prefactor $(4-n)$ 
and thus drop out in phenomenological applications. One 
only has to know box integrals.
\begin{eqnarray}\label{EQSR5}
I_5^n &=&  -\frac{1}{2}
\sum\limits_{l,k=1}^5 S^{-1}_{lk}  I_{4,l}^n  + {\cal O}(\epsilon)
\end{eqnarray}
Explicit expressions for the 5-point function with zero or one massive legs
may be found in \cite{BernDixonKosower,CampbellGloverMiller}. Other cases
are easily constructed by using known representations of
off-shell box-integrals.

\subsubsection*{Case N=6:}
Now the Gram matrix is not invertible anymore for 4-dimensional 
kinematics whereas an inverse for $S$ still exists.
With $(p_j+p_{j+1})^2=s_{j,j+1}$,
$(p_j+p_{j+1}+p_{j+2})^2=s_{j,j+1,j+2}$ $(j\mbox{ mod } 6)$ one has
\begin{eqnarray}\label{EQS6}
S = -\frac{1}{2}\,\left(  \begin{array}{cccccc} 
             0   &   m_2^2  &  s_{23}    & s_{234}    & s_{61}   & m_1^2  \\
          m_2^2  &    0     &  m_3^2  & s_{34}     & s_{345}  & s_{12}    \\
            s_{23}  &   m_3^2  &   0     &   m_4^2 & s_{45}   & s_{123}   \\
           s_{234}  &  s_{34}     &  m_4^2  &   0     & m_5^2 & s_{56}    \\
            s_{61}  &  s_{345}    &  s_{45}    &   m_5^2 & 0     & m_6^2  \\
          m_1^2  &  s_{12}     &  s_{123}   & s_{56}     & m_6^2 &  0    
\end{array}  \right)
\end{eqnarray}
Before going on, it is instructive to clarify the
dependences between the Mandelstam variables which define the matrix $S$. 
Since they are directly related to the cuts of the $N$-point graph, 
the following counting holds for $N\ge 4$, where 
 $M$ is the number of off-shell legs,
$C$ the number of cuts, $D$ the number of independent Lorentz
invariants built out of the vectors $r_1,\dots ,r_N$, and 
$Z=C-D$ the number of constraints:
\begin{eqnarray}
C=N(N-3)/2+M \; , \; D=3N-10+M \; , \; Z=(N-4)(N-5)/2 
\end{eqnarray}  
This makes manifest that for $N\ge 6$ one encounters subtleties due to 
the appearing constraints.
Here, the (nonlinear) constraint det$(G)=0$ shows up in the fact
that $\sum_{l=1}^N b_{l}=-\frac{1}{2}\sum_{k,l=1}^N S^{-1}_{kl}=0$.\\
The reduction formula for $N=6$ reads
\begin{eqnarray}\label{EQSR6}
I_6^n = -\frac{1}{2}
\sum\limits_{l,k=1}^6 S^{-1}_{lk}  I_{5,l}^n
\end{eqnarray}
If all six external legs are on-shell, the occurring pentagon integrals 
$I_{5,l}^n$
have one external leg off-shell. 
The explicit expression for the on-shell 6-point function is given in 
Appendix \ref{APPA}.

\section{Reduction of tensor-integrals}

In this section we show that any rank $L$, $N$-point tensor integral 
can, by recursion and scalar reduction, be expressed in terms of scalar 
box, triangle and two-point integrals. We prove that for $N\ge 5$ 
and non-exceptional external momenta higher dimensional integrals drop out.

\medskip

The rank $L$, $N$-point tensor integral is given by 
\begin{equation}
I_N^{{\mu_1}\dots {\mu_L}} = \int d\kappa\; 
\frac{k^{\mu_1}\dots k^{\mu_L}}{\prod^N_{l=1} q_l^2}
\label{EQdeft}
\end{equation}
We omit the superscript $n$ indicating the dimension in the tensor integrals 
to simplify the notation. If the dimension is different from 
$(4-2\epsilon)$, it will 
always be written out explicitly. \\
After introducing Feynman parameters and making a shift of the loop momentum
as in the last section, 
the odd powers of the loop momentum in the numerator can be dropped.
Using the standard integral
\begin{equation}
\int d\kappa \frac{k^{\mu_1}\dots k^{\mu_{2m}}}{(k^2-M^2)^N}
=(-1)^N \Bigl[ g^{\cdot\cdot}_{(m)} \Bigr]^{\{\mu_1\dots \mu_{2m}\}} 
\left(-\frac{1}{2}\right)^m
\frac{\Gamma(N-(n+2m)/2)}{\Gamma(N)} \left( M^2 \right)^{-N+(n+2m)/2}
\end{equation}
one finds the following formula:
\begin{equation}
I^{\mu_1 \dots \mu_L}_N = \sum \limits_{m=0}^{[L/2]} 
\left(-\frac{1}{2}\right)^m 
\sum \limits_{j_1,\dots,j_{L-2m}=1}^{N-1} 
\left[ g^{\cdot\cdot}_{(m)} r_{j_1}^{\cdot}\dots r_{j_{L-2m}}^{\cdot} \right]
^{\{\mu_1 \dots \mu_L\}}I_N^{n+2m}(j_1,\dots,j_{L-2m})
\label{EQdeftp}
\end{equation}
$[L/2]$ is the nearest integer less or equal to $L/2$ and 
$\left[ g^{\cdot\cdot}_{(m)} r_{j_1}^{\cdot}\dots r_{j_{L-2m}}^{\cdot} 
\right]^{\{\mu_1 \dots \mu_L\}}$
stands for the sum over all different combinations of $L$ Lorentz indices 
distributed to $m$ metric tensors and $(L-2m)$ $r$-vectors. 
These are ${L \choose 2m} \prod_{k=1}^m (2k-1)$ terms. 
A dot appearing as an index at objects inside a square bracket
stands for one index out of the set specified 
in curly brackets  at the outside  of the square bracket.
 $X^{(m)}$ or $X_{(m)}$ denotes the product of $m$ terms of $X$ with 
adequate indices\footnote{The ``power'' $(m)$ may appear as a lower index for convenience 
of notation if it would interfere with the dots standing for upper indices. 
Hence we write $X_{(m)}^{\cdot\cdot}$ instead of $(X^{\cdot\cdot})^{(m)}$.}. 
For example, a tensor object like 
$\left[ X_{(2)}^{\cdot\cdot} Y^{\cdot} \right]^{\{\mu_1 \mu_2 \mu_3\mu_4\mu_5\}}$
with $X$ a symmetric tensor of rank 2 and Y a vector means
\begin{displaymath}
\left[ X_{(2)}^{\cdot\cdot} Y^{\cdot} \right]^{\{\mu_1\mu_2\mu_3\mu_4\mu_5\}}
= \left(X^{\mu_1\mu_2}X^{\mu_3\mu_4} + X^{\mu_1\mu_3}X^{\mu_2\mu_4} +
X^{\mu_1\mu_4}X^{\mu_2\mu_3} \right)Y^{\mu_5}\; + 
5\;\mbox{permutations}.
\end{displaymath}
$I_N^D(j_1,\dots ,j_M)$ are scalar  
integrals with nontrivial numerators in $D$ dimensions, defined by
\begin{eqnarray}\label{EQFeynparint}
I_N^D(j_1,\dots ,j_M) &=& (-1)^N \Gamma (N-D/2) 
\int_{0}^{\infty} d^Nz \,\delta(1-\sum\limits_{l=1}^N z_{l})
\frac{z_{j_1}\dots z_{j_M}}{(z\cdot S\cdot z)^{N-D/2}} 
\end{eqnarray}
Recursion relations for this kind of integrals
were given for $M\le 4$ in \cite{CampbellGloverMiller}.
We derive such relations for general $M$ in Appendix \ref{APPB}.

From now on we  use standard summation conventions for the indices. 
Eq.~(\ref{EQdeftp}) implies that 
tensor integrals in momentum space are linear combinations
of scalar integrals in different dimensions.
We note that formula (\ref{EQdeftp}) is equivalent 
to a more general formula given in \cite{Davydychev} for the case of 
integer powers of the propagators.

\medskip

We want to prove now that for non-exceptional momenta and $N\ge 5$ 
the higher dimensional
scalar integrals drop out. To this end we solve Eq.~(\ref{EQdeftp}) for
$I_N^{n}(j_1,\dots,j_L)$ by contracting it with 
$2^L r^{\mu_1}_{l_1}\dots r^{\mu_L}_{l_L}$.
For the inversion of the Gram matrices we use its (pseudo) inverse as defined 
above.
With $2 r_l\cdot k = q_N^2 - q_l^2 + v_l$ one finds
\begin{eqnarray}\label{EQscalarL}
I_N^n(j_1,\dots,j_L) &=& \sum\limits_{k=0}^{L} \Bigl[ w^{(L-k)}_{\cdot}
H_{\cdot l_1} \dots H_{\cdot l_k} \Bigr]_{\{j_1,\dots,j_L\}} 
I^n_{N-k,N-l_1,\dots,N-l_k}\nonumber\\&&
 - \sum \limits_{m=1}^{[L/2]} \left(-1\right)^m 
\left[ H^{(m)}_{\cdot\cdot}  I_N^{n+2m}(\mbox{$(L-2m)$ indices} )
\right]_{\{j_1,\dots,j_L\}}
\end{eqnarray}
where $w=H\cdot v$ and $H$ is defined in Eq.~(\ref{EQdefpseudoinvers}). 
On the right-hand side of Eq.~(\ref{EQscalarL}),  
differences of reduced integrals appear as defined in 
Eq.~(\ref{EQreducedgraph}).
In principle one has to add to (\ref{EQscalarL}) also 
the solutions stemming from the homogeneous equation 
which are present if $G$ is not of maximal rank.
But as they contain the matrix $K=1_{N-1}-H\, G$,
they vanish after contraction with the $r$'s, i.e. 
$r^{\mu}\cdot (1_{N-1}-H\, G) =0$. If $G$ is invertible, $H=G^{-1}$.
The bracket of the $k$-th term of the first sum stands for $ L \choose k $
terms whereas the bracket of the $m$-th term of the second sum 
stands for ${L \choose 2m} \prod_{k=1}^m (2k-1)$ terms.
Inserting (\ref{EQscalarL}) into (\ref{EQdeftp}), we find that 
the tensor integral decays into a part containing 
$(4-2\epsilon)$-dimensional objects, $K_N^{\mu_1 \dots \mu_L}$, and a part
built out of  higher-dimensional objects, $J_N^{\mu_1 \dots \mu_L}$.
\begin{eqnarray}
I^{\mu_1 \dots \mu_L}_N &=& K^{\mu_1 \dots \mu_L}_N + 
J^{\mu_1 \dots \mu_L}_N \label{EQgeneraltred}\\
K^{\mu_1 \dots \mu_L}_N &=& \sum\limits_{k=0}^{L} \Bigl[ {\cal W}_{(L-k)}^{\cdot}
{\cal K}_{l_1}^{\cdot} \dots {\cal K}_{l_k}^{\cdot} 
\Bigr]^{\{\mu_1,\dots,\mu_L\}} 
I^n_{N-k,N-l_1,\dots,N-l_k}  \label{EQgeneraltred0}\\
J^{\mu_1 \dots \mu_L}_N  &=&  \sum \limits_{m=1}^{[L/2]} (-1)^m 
\sum \limits_{j_1,\dots,j_{L-2m}=1}^{N-1} 
\left[ \Bigl((g/2)_{(m)}^{\cdot\cdot}-{\cal H}_{(m)}^{\cdot\cdot}\Bigr) 
r_{j_1}^{\cdot}\dots r_{j_{L-2m}}^{\cdot} 
\right]^{\{\mu_1 \dots \mu_L\}}\nonumber\\
&&\times \,I_N^{n+2m}(j_1,\dots,j_{L-2m}) \label{36b}
\end{eqnarray}
The Lorentz indices are carried by the objects
\begin{eqnarray}
{\cal H}^{\mu\nu} &=&  r^\mu \cdot H \cdot r^\nu  \nonumber \\
{\cal K}^{\mu}_l &=&  ( r^\mu \cdot H )_l \nonumber \\
{\cal W}^\mu &=& r^\mu\cdot w = {\cal K}^{\mu}\cdot v \label{Lorentz}
\end{eqnarray}
We recall that $w=H\cdot v$ and $v_l=G_{ll}/2$.

If the $r$'s span 4-dimensional Minkowski space,
which is generically the case if $N\ge 5$, ${\cal H}^{\mu\nu}$ 
is just proportional to  the metric tensor $g_{4}^{\mu\nu}$ in 4 dimensions:
\begin{eqnarray}\label{EQrHr}
{\cal H}^{\mu\nu} = r^{\mu}\cdot H\cdot r^{\nu} =
\sum\limits_{i,j=1}^{N-1} r^{\mu}_i r^{\nu}_j H_{ij} = 
\sum\limits_{l,k=1}^{4} E^{\mu}_l E^{\nu}_k \tilde G_{lk}^{-1} 
= \frac{1}{2} g_{4}^{\mu\nu}
\end{eqnarray}  
Thus the coefficients of the the higher dimensional integrals in (\ref{36b}) 
are proportional to $g^{(m)}-g^{(m)}_4$ which is of order $\epsilon$.
As by power counting the higher dimensional integrals are
finite objects if the external momenta are 4-dimensional, 
the whole contribution is $\cal{O}(\epsilon)$. 
This proves the cancellation of higher dimensional
integrals in tensor reductions for arbitrary $N\ge5$ and non-exceptional 
external momenta, and we obtain  
\begin{eqnarray}
I^{\mu_1 \dots \mu_L}_N &=& \sum\limits_{k=0}^{L} \Bigl[ {\cal W}_{(L-k)}^{\cdot} 
{\cal K}_{l_1}^{\cdot} \dots {\cal K}_{l_k}^{\cdot} 
\Bigr]^{\{\mu_1,\dots,\mu_L\}} 
I^n_{N-k,N-l_1,\dots,N-l_k} + {\cal O}(\epsilon) \quad ,\quad N\ge5
\label{EQtreduction}
\end{eqnarray}
For $N<5$, Eq.~(\ref{EQrHr}) is not valid since the external 
momenta cannot  span Minkowski space anymore. Thus one has to calculate the terms 
$J^{\mu_1 \dots \mu_L}_N$ for $N<5$.  The explicit expressions 
are given in Appendix \ref{appexplicit}.

\medskip

Now we want to rewrite Eq.~(\ref{EQgeneraltred}) as a recursion formula for 
arbitrary tensor integrals. To this effect we express the contracted tensor
integrals $I_{N-k,N-l_1,\ldots,N-l_k}^n$ $(k>0)$ as 
$(N-1)$-point tensor integrals which are maximally  of rank $(L-1)$.
By using $(k-1)$ times the relation $q_N^2-q_l^2=2\,r_l\cdot k-v_l$, one gets
\begin{eqnarray}
I_{N-k,N-l_1,\ldots,N-l_k}^n
&=& \sum\limits_{j=0}^{k-1} (-1)^{k-j-1}\frac{2^{j}}{j!}
\Bigl[v^{(k-j-1)}_{\cdot}\,
r_{\cdot\nu_1}\ldots
r_{\cdot\nu_{j}}\Bigr]_{\{l_1\ldots l_{k-1}\}}
I_{N-1,N-l_k}^{\nu_1\ldots\nu_{j}}\; , \; k\ge 1
\label{Ik}
\end{eqnarray}
Insertion of  expression (\ref{Ik}) into Eq.~(\ref{EQgeneraltred0}) leads to the 
recursion formula 
\begin{eqnarray}\label{rec}
K_N^{\mu_1\ldots\mu_L}&=&\frac{1}{L} \Bigl[{\cal W}^{\cdot} 
K_N^{\{L-1 \;\mathrm{dots}\}}\Bigr]^{\{\mu_1\ldots\mu_L\}}\nonumber\\
&&+\frac{2^{(L-1)}}{L!}
\Bigl[{\cal K}_l^{\cdot}{\cal H}^{\cdot}_{\nu_1}
\ldots{\cal H}^{\cdot}_{\nu_{L-1}}\Bigr]^{\{\mu_1\ldots\mu_L\}}
I_{N-1,N-l}^{\nu_1\ldots\nu_{L-1}}
\end{eqnarray}
The derivation of (\ref{rec}) is given in Appendix \ref{apprec}. \\
Since for $N\ge 5$ the terms $J^{\mu_1 \dots \mu_L}_N$ contribute only to 
order $\epsilon$, $K_N^{\mu_1 \dots \mu_L}$ in Eq.~(\ref{rec}) can be replaced by 
$I_N^{\mu_1 \dots \mu_L}$ to obtain the recursion formula for $N\ge 5$. 
For $N<5$, one cannot drop the $J_N^{\mu_1\ldots\mu_L}$ terms,
 such that the 
general recursion formula reads
\begin{eqnarray}
I_N^{\mu_1\ldots\mu_L}&=& J_N^{\mu_1\ldots\mu_L} + \frac{1}{L}
\Bigl[{\cal W}^{\cdot}(I_N-J_N)^{\{L-1 \;\mathrm{dots}\}}
\Bigr]^{\{\mu_1\ldots\mu_L\}}\nonumber\\
&&+\frac{2^{(L-1)}}{L!}
\Bigl[{\cal K}_l^{\cdot}{\cal H}^{\cdot}_{\nu_1}\ldots{\cal H}^{\cdot}_{\nu_{L-1}}
\Bigr]^{\{\mu_1\ldots\mu_L\}}I_{N-1,N-l}^{\nu_1\ldots\nu_{L-1}}
\label{recwithJ}
\end{eqnarray}
As a consequence, 
higher dimensional integrals $I_{N^{\prime}}^{n+2m}\,(m=1,2)$ will 
appear in the reduction of tensor $N$-point integrals only with 
$N^{\prime}\le 4$. All the higher dimensional integrals can be 
re-mapped to $n$-dimensional integrals with the scalar reduction formulas 
(\ref{EQSR2}), (\ref{EQSR3}) and (\ref{EQSR4}). \\
In the case of exceptional kinematics and $N\ge 5$, 
higher dimensional $(N\ge 5)$-point functions can be present.  

\medskip

As some of the reduced integrals on the right-hand side 
of (\ref{recwithJ}) do not contain 
a trivial propagator, they are not in the standard form for applying 
the reduction formula again. Therefore one has to perform  a shift 
$k\rightarrow k+r_{l}$ in the tensor integral, leading to
\begin{eqnarray}\label{EQshift}
I^{\mu_1\dots\mu_P}_{N-1,N-l}(R)
&=&  I^{\mu_1\dots\mu_P}_{N-1}(\hat R_{[N]})-
I^{\mu_1\dots\mu_P}_{N-1}(\hat R_{[l]})\nonumber\\
&=& \sum\limits_{k=0}^{P} \Bigl[ r_{l}^{\cdot (k)} 
I_{N-1}^{\{P-k \;\mathrm{dots}\}}(R_{[l]})\Bigr]^{\{\mu_1\dots\mu_{P}\}}-
I_{N-1}^{\mu_1\dots\mu_{P}}(\hat R_{[l]})
 \end{eqnarray}
  
The argument vectors are
\begin{eqnarray}\label{EQargumentvectors}
R &=& (r_1,\dots,r_{N}) \nonumber\\
\hat R_{[k]} &=& (r_{1},\dots,\hat r_{k} ,\dots , r_N)\nonumber\\
R_{[l]}   &=& (r_{l+1}-r_{l},r_{l+2}-r_{l},\dots,r_{N-2+l}-r_{l},0)
\end{eqnarray}
where $\hat r$ means that the respective vector has to be left out.
The vector indices are understood to be taken cyclically symmetric
with periodicity $N$, i.e.
$r_{N+l}=r_{l}$ for $l\in \{1,\dots,N\}$.
As we assume $r_N=0$, the integrals in the last line of (\ref{EQshift})
have again at least one trivial propagator and are 
suited for a further reduction step.

A full tensor reduction of a rank $L$ tensor integral 
is obtained by employing Eq.~(\ref{rec}) 
for $N\ge 5$ resp. Eq.~(\ref{recwithJ}) for $N<5$ 
to do the first reduction step. 
Then the integrals $I_{N-1,N-l}^{\nu_1\ldots\nu_{L-1}}$ 
have to be shifted by using Eq.~(\ref{EQshift}) to end up with 
expressions of  rank $\le L-1$ to which the same reduction procedure 
can be applied again. 
Explicit tensor reduction formulas for $N=2,\ldots,6$ are given in 
Appendix \ref{appexplicit}. 

For $N\ge 5$ one can express by recursion any rank $L$, $N$-point tensor 
integral in terms of $(4-2\epsilon)$-dimensional scalar $N$-point and 
tensor box integrals, 
which in turn can be reduced further down to scalar box, 
triangle and two-point integrals, taking also into account the scalar 
reduction formulas given in the previous section. 
This defines an algorithm which can be easily programmed.

\section{Conclusion}

We have presented reduction formulas for massless 
$N$-point scalar and tensor functions.
 
We have shown in a constructive manner how scalar $N$-point functions 
can be represented as linear combinations of 
$(4-2\epsilon)$-dimensional scalar $(N-1)$-point 
functions for arbitrary $N\ge 5$.
In particular, we pointed out how to treat the linear
equations which determine the reduction coefficients, $b_l$, for $N\ge 6$ 
in terms of the pseudo-inverse of the singular Gram matrix. 
In this way all mathematical operations are valid
for 4-dimensional external kinematics throughout the whole reduction 
procedure.\\
We applied the reduction formalism to scalar integrals
up to $N=6$ explicitly and gave an expression 
for the on-shell 6-point function.

For $N$-point tensor integrals we have formulated a reduction scheme for 
arbitrary $N$. We have proven that
higher-dimensional integrals always vanish  for $N\ge 5$ in the case of
non-exceptional kinematics. We  derived a recursion formula for
the remaining $n$-dimensional part. 
For the derivation we used methods \`a la Passarino-Veltman,
 such as contracting tensor integrals with external vectors and inverting 
Gram matrices. In the case of a singular Gram matrix the inversion 
has to be done  with its pseudo-inverse.
By iteration any tensor integral 
can be expressed as a linear combination of 
scalar integrals. Higher dimensional scalar integrals
appear only in the reduction of $N\le 4$ tensor integrals.
These higher dimensional integrals are expressible
in terms of $(4-2\epsilon)$-dimensional integrals as has been shown  explicitly
in the discussion of the scalar reduction formulas.
To make the general formalism more user-friendly
we gave explicit expressions up to $N=6$
in an appendix.

We also derived  a reduction formula in Feynman
parameter space in an appendix. 
The formula generalizes results in the literature
avoiding a projective transformation. 
As we gave all the formulas to translate  momentum space
expressions into Feynman parameter space expressions,  
the equivalence between the two approaches
is manifest. In applications this will allow us to employ
methods similar to the ones in \cite{CampbellGloverMiller} to get
numerically stable expressions, a problem we did
not address in this paper.
  
We conclude that the computation of IR divergent
one-loop integrals for arbitrary numbers of legs
can be mastered with the reduction formulas presented here.
The iterative structure makes it easy to
implement the formalism in algebraic computer programs.
The conceptual problems 
for the construction of multi-parton one-loop amplitudes
are thus solved. 

The generalization of our method to include also massive particles  
is postponed to a future publication.   
 
\vspace{1.5cm}

{\bf\Large Acknowledgements}

\vspace{5mm}

We thank Bas Tausk and Christian Schubert for 
interesting remarks and discussions. 
G.H. would like to thank the LAPP for its hospitality during her stay 
in June 1999. \\
This work was supported in part by the EU Fourth Training Programme  
''Training and Mobility of Researchers'', Network ''Quantum Chromodynamics
and the Deep Structure of Elementary Particles'',
contract FMRX--CT98--0194 (DG 12 - MIHT).

\vspace{1cm}

\appendix

\section{The scalar 6-point function}\label{APPA}
Here we present an explicit formula for the scalar hexagon function
with all external legs on-shell. The reduction formula (\ref{EQSR6}) 
allows to write it as a sum of six pentagon integrals with
one external leg off-shell.
\begin{eqnarray}
&&I_6^n(s_{12},s_{23},s_{34},s_{45},s_{56},s_{61},s_{123},s_{234},s_{345}) = \vspace{5cm}\nonumber\\
&&-\frac{1}{2}\sum\limits_{k,l=1}^{6} S^{-1}_{kl}
I_{5,1mass}(s_{l+2,l+3},s_{l+3,l+4},s_{l+4,l+5},s_{l+5,l,l+1},s_{l,l+1,l+2};s_{l,l+1})
\label{In6}
\end{eqnarray}
The function $I_{5,1mass}$ has been given in \cite{BernDixonKosower}.
We rederived the formula. Although we agree with it 
for Euclidean
kinematics we note that the expression given there is not well suited for
analytic continuation\footnote{The problematic term (up to a trivial typo) 
is the bracket $(1-(m_5^2 s_{23}/s_{45}/s_{51})^{-\epsilon})$. 
Rederiving the formula we actually
get  $(1- (m_5^2/s_{45})^{-\epsilon} (s_{23}/s_{51})^{-\epsilon})$.
Only in the latter form the analytic continuation is defined by simply adding an
infinitesimal imaginary part to the Mandelstam variables. Care has to be taken
as well in the analytic continuation of $Li_2(1-m_5^2 s_{23}/s_{45}/s_{51})$,
cf. the remarks at the end of this Appendix.}.
The other ingredient, the matrix
$S$, is given explicitly 
by Eq.~(\ref{EQS6}) after setting all $m_j^2=0$.
The indices in (\ref{In6}) have to be taken modulo 6, and due to
 momentum conservation
one has $s_{456}=s_{123}$, $s_{612}=s_{345}$, $s_{561}=s_{234}$.
The nine Lorentz invariants are not independent since they fulfill the 
constraint $\det(G)=0$.
We find
\begin{eqnarray}
I_6^n &=& \sum\limits_{k=1}^{6} \Bigl[\,
\frac{r_\Gamma}{\epsilon^2} ( A_k + B_k ) + C_k + D_{k}  \Bigr]
	   \\ 
A_1 &=&  \frac{(-s_{12})^{-\epsilon}}{s_{61}s_{12}s_{23}s_{234}} \nonumber\\
B_1 &=& \frac{(s_{12}-s_{123})^2 [(-s_{12})^{-\epsilon}-(-s_{123})^{-\epsilon}]}
            {s_{12}s_{23}s_{34}s_{123} E_1}
       +\frac{(s_{12}-s_{345})^2 [(-s_{12})^{-\epsilon}-(-s_{345})^{-\epsilon}]}
            {s_{56}s_{61}s_{12}s_{345} E_1}  \nonumber\\
C_1 &=& b_1 \left( \frac{1}{s_{34}s_{45}s_{56}} + 
                   \frac{s_{45}-s_{345}}{s_{34}s_{45}E_1} +
		   \frac{s_{45}-s_{123}}{s_{45}s_{56}E_1}\right)
		   \mbox{Li}_2\left(1-\frac{s_{123}s_{345}}{s_{12}s_{45}}\right)
	- \frac{b_1}{s_{34}s_{45}s_{56}}\frac{\pi^2}{3}\nonumber\\&&
\qquad\quad 
    +  \frac{(s_{123}-s_{12})^2}{s_{12}s_{23}s_{34}s_{123}E_1}
	        \left[\mbox{Li}_2\left(1-\frac{s_{12} }{s_{123}}\right) -
	              \mbox{Li}_2\left(1-\frac{s_{123}}{s_{12} }\right)	\right]   
\nonumber\\&&
\qquad\quad 
    +  \frac{(s_{345}-s_{12})^2}{s_{56}s_{61}s_{12}s_{345}E_1}
	        \left[\mbox{Li}_2\left(1-\frac{s_{12} }{s_{345}}\right) -
	              \mbox{Li}_2\left(1-\frac{s_{345}}{s_{12} }\right)	\right] 
\nonumber\\
D_{1} &=&  -\frac{1}{2}
         \left[ b_4 
      \left(  \frac{1}{s_{61}s_{12}s_{23}} + 
              \frac{s_{12}-s_{123}}{s_{12}s_{23}E_1} -
              \frac{s_{12}-s_{345}}{s_{61}s_{12}E_1} \right) \right.
\nonumber\\&&  \qquad
        \left. +b_3 
      \left(  \frac{1}{s_{56}s_{61}s_{12}} + 
              \frac{s_{61}-s_{234}}{s_{56}s_{61}E_3} -
              \frac{s_{61}-s_{345}}{s_{61}s_{12}E_3} \right) \right]
	   \log^2\left( \frac{s_{12}}{s_{61}}\right)   
\nonumber\\&&-\frac{1}{2}
         \left[ b_3 
      \left(  \frac{1}{s_{56}s_{61}s_{12}} - 
              \frac{1}{s_{56}s_{61}s_{345}} -
              \frac{1}{s_{61}s_{12}s_{234}} \right) \right.
\nonumber\\&&  \qquad
        \left. +b_1
      \left(  \frac{s_{12}-s_{345}}{s_{56}s_{345}E_1} -
              \frac{s_{12}-s_{123}}{s_{34}s_{123}E_1} \right)
	       +b_5
      \left(  \frac{s_{56}-s_{234}}{s_{12}s_{234}E_2} -
              \frac{s_{56}-s_{123}}{s_{34}s_{123}E_2} \right)      
	      \right]
	   \log^2\left( \frac{s_{12}}{s_{56}}\right) 
\nonumber\\&& -\frac{1}{2}
         \left[	  b_2
      \left(  \frac{1}{s_{45}s_{56}s_{61}} + 
              \frac{s_{56}-s_{123}}{s_{45}s_{56}E_2} +
              \frac{s_{56}-s_{234}}{s_{56}s_{61}E_2} \right) \right.
\nonumber\\&&  \qquad
        \left. +b_5 
      \left(  \frac{1}{s_{12}s_{23}s_{34}} + 
              \frac{s_{23}-s_{123}}{s_{12}s_{23}E_2} +
              \frac{s_{23}-s_{234}}{s_{23}s_{34}E_2} \right) \right]
	   \log^2\left( \frac{s_{123}}{s_{234}}\right) 
\nonumber\\&& -\frac{1}{2}
         \left[	  b_1
      \left(  \frac{s_{12}-s_{123}}{s_{34}s_{123}E_1} \right) 
           -b_4 
      \left(  \frac{1}{s_{61}s_{12}s_{23}} + 
              \frac{s_{12}-s_{345}}{s_{61}s_{12}E_1} \right) \right.
\nonumber\\&&  \qquad\qquad\qquad\qquad\qquad
        \left. +b_5 
      \left(  \frac{s_{56}-s_{123}}{s_{34}s_{123}E_2} +
              \frac{s_{23}-s_{123}}{s_{12}s_{23}E_2} \right) \right]
	   \log^2\left( \frac{s_{12}}{s_{123}}\right) 
\nonumber\\&& -\frac{1}{2}
         \left[	  b_3 
      \left(  \frac{1}{s_{61}s_{12}s_{234}} - \frac{1}{s_{56}s_{61}s_{12}} -
              \frac{s_{61}-s_{234}}{s_{56}s_{61}E_3} -
              \frac{s_{34}-s_{234}}{s_{12}s_{234}E_3} \right) \right.
\nonumber\\&&  \qquad
        \left. +b_5 
      \left(  \frac{1}{s_{12}s_{23}s_{234}} - \frac{1}{s_{12}s_{23}s_{34}} -
              \frac{s_{23}-s_{234}}{s_{23}s_{34}E_2} -
              \frac{s_{56}-s_{234}}{s_{12}s_{234}E_2} \right) \right]
	   \log^2\left( \frac{s_{12}}{s_{234}}\right) 
\nonumber\\&& -\frac{1}{2}
         \left[	  b_1
      \left(  \frac{s_{12}-s_{345}}{s_{56}s_{345}E_1} \right) 
           -b_4
      \left(  \frac{1}{s_{61}s_{12}s_{23}} + 
              \frac{s_{12}-s_{123}}{s_{12}s_{23}E_1} \right) \right.
\nonumber\\&&  \qquad\qquad\qquad\qquad\qquad
        \left. +b_3
      \left(  \frac{s_{34}-s_{345}}{s_{56}s_{345}E_3} +
              \frac{s_{61}-s_{345}}{s_{61}s_{12}E_3} \right) \right]
	   \log^2\left( \frac{s_{12}}{s_{345}}\right) \nonumber
\end{eqnarray}
with the abbreviations
\begin{eqnarray}
E_1 &=& E_4 = s_{123} s_{345} - s_{12} s_{45} \nonumber\\
E_2 &=& E_5 = s_{234} s_{123} - s_{23} s_{56} \nonumber\\
E_3 &=& E_6 = s_{345} s_{234} - s_{34} s_{61} \nonumber
\end{eqnarray}
and
$$r_{\Gamma}=\frac{\Gamma(1+\epsilon)\Gamma^2(1-\epsilon)}{\Gamma(1-2\epsilon)}$$
The functions $A_{k},B_{k},C_{k},D_{k}$ for $k>1$ are obtained by
cyclic permutations of the indices. The other necessary
ingredients are
\begin{eqnarray}
b_1  &=& (-s_{345}^2 s_{56}^2 s_{23}+s_{12}s_{234}s_{45}^2s_{34}
-s_{56}s_{23}s_{345}s_{45}s_{34}-s_{45}s_{234}s_{345}s_{123}s_{34} \nonumber\\&&
+s_{34}^2s_{45}s_{61}s_{123}-s_{34}s_{123}s_{12}s_{234}s_{45}
-s_{34}s_{123}s_{56}s_{23}s_{345}+2s_{34}s_{45}s_{23}s_{12}s_{56}\nonumber\\&&
+s_{34}s_{123}^2s_{345}s_{234}-s_{34}^2s_{61}s_{123}^2
-s_{34}s_{345}s_{56}s_{123}s_{61}+s_{345}^2s_{56}s_{123}s_{234}\nonumber\\&&
-s_{45}s_{234}s_{123}s_{56}s_{345}+s_{45}^2s_{234}s_{12}s_{56}
+s_{56}^2s_{23}s_{345}s_{45}-s_{56}s_{34}s_{45}s_{61}s_{123}\nonumber\\&&
+2s_{34}s_{12}s_{56}s_{61}s_{45}-s_{12}s_{56}s_{45}s_{234}s_{345}
-s_{12}^2s_{234}s_{45}^2-s_{123}^2s_{234}s_{345}^2\nonumber\\&&
+s_{123}s_{56}s_{23}s_{345}^2+2s_{45}s_{345}s_{234}s_{123}s_{12}
-2s_{34}s_{45}^2s_{12}s_{56}+2s_{45}s_{34}s_{123}s_{56}s_{345}\nonumber\\&&
-s_{45}s_{34}s_{61}s_{123}s_{12}-s_{45}s_{56}s_{23}s_{345}s_{12}
+s_{34}s_{123}^2s_{61}s_{345})/F
\end{eqnarray}
where the other $b_k$'s are again obtained by cyclic permutation. Finally,
\begin{eqnarray}
F &=&  -s_{345}^2s_{56}^2s_{23}^2+4s_{23}s_{34}s_{12}s_{56}s_{61}s_{45}
-2s_{23}s_{34}s_{345}s_{56}s_{123}s_{61}
-2s_{23}s_{12}s_{56}s_{45}s_{234}s_{345}\nonumber\\&&
+2s_{23}s_{345}^2s_{56}s_{123}s_{234}
-s_{123}^2s_{61}^2s_{34}^2-2s_{34}s_{12}s_{123}s_{234}s_{61}s_{45}
+2s_{34}s_{123}^2s_{61}s_{345}s_{234}\nonumber\\&&
-s_{45}^2s_{234}^2s_{12}^2
-s_{345}^2s_{123}^2s_{234}^2+2s_{12}s_{123}s_{234}^2s_{45}s_{345} \nonumber\\
&=& 64\,\det(S) 
\end{eqnarray} 
Note that in the $\epsilon$-dependent part of the formula
no $b$'s appear. We did not succeed in finding a more compact form
for the part which does not depend on $\epsilon$ 
without spoiling this nice feature. 

The expression for $I_6^n$ in the form as given above is strictly only correct in the
Euclidean region where all Mandelstam variables are 
negative. For most of the terms, the analytic continuation to 
positive values is defined 
by simply using the replacement $s\rightarrow s+i\delta$, where
$s$ stands here for any of the $s_{ij}$, $s_{ijk}$.
No cut will be hit by the logarithms, the dilogarithms
with a single ratio of Mandelstam variables and
the exponentials $(-s-i\delta)^{-\epsilon}$.
Concerning the dilogarithms of a product of ratios, more care
has to be taken. To avoid the crossing of a cut
one has to make the replacement
\begin{eqnarray}
Li_2\left(1-\frac{s_1s_2}{s_3s_4}\right) 
&\rightarrow& Li_2\left(1-\frac{s_1+i\delta}{s_3+i\delta}
          \frac{s_2+i\delta}{s_4+i\delta}\right) \nonumber\\
	  &&+\; \eta\left(\frac{s_1+i\delta}{s_3+i\delta} ,
	              \frac{s_2+i\delta}{s_4+i\delta}\right) 
		      \log\left(1-\frac{s_1+i\delta}{s_3+i\delta}
          \frac{s_2+i\delta}{s_4+i\delta}\right) \\
\eta(x,y) &=& \log(xy) - \log(x) - \log(y)
\end{eqnarray}

\section{Recursion relation for integrals with Feynman parameters in the 
numerator}\label{APPB}
Here we want to comment on the reduction of Feynman parameter
integrals with nontrivial numerators as defined in (\ref{EQFeynparint}).
In \cite{CampbellGloverMiller} recursion relations
for integrals with up to 4 Feynman parameters are given.
The derivation is based on the approach of \cite{BernDixonKosower}
which uses a projective transformation \cite{tHooftVeltman}. 
We present an independent derivation and generalize the
formulas to arbitrary $N$ and numbers of Feynman parameters in the
numerator.

Consider the following identity ($j=1,\dots,N-1$)
\begin{equation}
\int\limits_{-\infty}^{\infty}  d^Nz   \frac{\partial}{\partial z_j} 
\left( \prod_{l=1}^{N} \theta(z_l) 
\delta(1-\sum_{k=1}^{N} z_k)  
z_{l_1} \dots z_{l_p}( z\cdot S\cdot z)^{-N+n/2+1} \right)=0
\end{equation}
Setting $r_N=0$, one obtains $z\cdot S\cdot z=\sum_{k,l=1}^{N-1}z_l G_{lk}z_k/2-
\sum_{k=1}^{N-1} v_{k} z_k$. The first step consists in eliminating
 the $\delta$-function by integrating out $z_N$. 
If now the derivative acts on $\theta$-functions, 
terms with $\delta$-function insertions are produced. 
The reduced integrals $I^n_{N-1,j}$ (defined in Eq.~(\ref{EQpinch}) below),
 obtained by pinching the $j$-th 
propagator line in
an $N$-point  graph, 
correspond to these $\delta$-function insertions.
If the derivative acts on the Feynman parameters, the monome in the
numerator  is reduced by one degree, 
and if it acts on the $z\cdot S\cdot z$
term it formally decreases the dimension by two and  increases the
numerator by $(G\cdot z)_j-v_j$. After these operations
one reintroduces the $z_N$-integration with a delta-function 
insertion.   
Using the following generalized definition
for pinched scalar integrals with nontrivial numerators 
\begin{eqnarray}\label{EQpinch}
I^n_{N-1,j}(l_1,\dots ,l_p) = (-1)^{N-1} \Gamma(N-1-n/2) 
\int_0^\infty d^Nz \delta(1-\sum_{l=1}^N z_l) 
\frac{z_{l_1}\dots z_{l_p}\delta(z_j)}{(z\cdot S\cdot z)^{N-1-n/2}}
\end{eqnarray}
we find
\begin{eqnarray}\label{EQreduction1}
\sum_{l_0=1}^{N-1} G_{j l_0} I_N^n(l_0,\dots ,l_p)=\sum\limits_{k=1}^p
\delta_{j l_k} I^{n+2}_N(l_1,\dots ,\hat l_k, \dots l_p)
+ I_{N-1,N}^n(l_1,\dots , l_p) \nonumber \\
- I_{N-1,j}^n(l_1 ,\dots ,l_p) + v_{j} I_N^n(l_1, \dots ,l_p)
\end{eqnarray}
Herein $\hat l_k$ means that the respective index does not appear.

Now we want to sketch the proof for a second equation
\begin{eqnarray}\label{EQreduction2}
I^n_{N-1,N}(l_1 ,\dots , l_p) = (N-n-p-1) I_N^{n+2}(l_1,\dots ,l_p)
+ \sum_{l_0=1}^{N-1} v_{l_0} I_N^{n}(l_0, \dots , l_p)
\end{eqnarray}
The proof is done by induction to $p$.
In order to show that the induction start $(p=0)$ holds true,
one directly calculates
  \begin{eqnarray}
I^n_{N-1,N}(1) = \int d\kappa \frac{k^2}{\prod_{l=1}^N q_l^2} 
=(N-n-1) I_N^{n+2}(1) + \sum_{l_0=1}^{N-1} v_{l_0} I_N^{n}(l_0)
\end{eqnarray}
For the induction step, one assumes that (\ref{EQreduction2})  is
fulfilled for a given $p$. Viewing the $v_j$
as independent variables, one differentiates  the formula
 with respect to $v_{l_{p+1}}$ and
finds the formula for $p+1$ and $n-2$.
As the dimension is arbitrary we can replace $n-2$
by $n$ in the expression. This proves the validity of equation 
(\ref{EQreduction2}) for all $p$.

Combining (\ref{EQreduction1}) and (\ref{EQreduction2})
one finds ($j=1,\dots, N$)
\begin{eqnarray}\label{EQreduction3}
2 \sum_{l_0=1}^{N} S_{jl_0} I_N^n(l_0,\dots ,l_p)=\sum\limits_{k=1}^p
\delta_{j l_k} I^{n+2}_N(l_1,\dots ,\hat l_k, \dots ,l_p)\hspace{4cm}\\
+ (N-n-p-1) I_N^{n+2}(l_1,\dots ,l_p) \nonumber
- I_{N-1,j}^n(l_1, \dots ,l_p)
\end{eqnarray}
In the case $N\le 6$, $S$ is invertible and the final reduction formula reads
\begin{eqnarray}\label{EQreduction4}
I_N^n(l_0,\dots ,l_p)=\frac{1}{2}\sum\limits_{k=1}^p
S^{-1}_{l_0 l_k} I^{n+2}_N(l_1,\dots ,\hat l_k, \dots ,l_p)\hspace{5cm}\\
+\frac{1}{2} \sum\limits_{j=1}^N S^{-1}_{j l_0}
 (N-n-p-1) I_N^{n+2}(l_1,\dots ,l_p) 
 -  \frac{1}{2} \sum\limits_{j=1}^N 
 S^{-1}_{j l_0}I_{N-1,j}^n(l_1,\dots ,l_p)\nonumber
 \end{eqnarray}
This is the generalization of the formulas given
in \cite{CampbellGloverMiller} for the case of four
Feynman parameters in the numerator $(p=3)$.
(In their conventions $S^{-1}_{ik}=\eta_{ik}\alpha_{i} \alpha_{k} /N_{n}$
and $\sum_i S^{-1}_{ik}=\gamma_{k}/N_{n}$, which is identical
to our $S^{-1}$ up to a trivial relabeling of indices).
It means that all Feynman parameter integrals with numerators
can be reduced to ordinary scalar integrals by iteration. 

%

In the case $N>6$, $S$ is not invertible. The linear dependence
of the row (resp. column) vectors of $S$ makes it difficult to
generalize the Feynman parameter space based techniques  to arbitrary
$N$~\cite{BernDixonKosower,FleischerJegerlehnerTarasov}.
In our approach  one can use Eq.~(\ref{EQreduction3}) 
directly in this case, where
the inversion of the Gram matrix should be done with 
its pseudo-inverse as explained in the main text.
As already noted earlier, it is also
possible to work with the pseudo-inverse to $S$. 
In any way, as we have proven, 
it is the (pseudo) inverse to the Gram matrix
which induces the cancellation of the higher dimensional terms 
by virtue of Eq.~(\ref{EQrHr}).

Finally, with the formulas given above 
and in the main text, it is possible to translate
expressions in Feynman parameter space into expressions
in momentum space and vice versa for arbitrary $N$.  


\section{Derivation of the recursion formula for tensor reduction}\label{apprec}

In this appendix we give the derivation of the recursion relation ~(\ref{rec}):
\begin{eqnarray}\label{recAP}
K_N^{\mu_1\ldots\mu_L}&=&\frac{1}{L}\Bigl[{\cal W}^{\cdot} 
K_N^{\{L-1 \,\mathrm{dots}\}}
\Bigr]^{\{\mu_1\ldots\mu_L\}}
+\frac{2^{(L-1)}}{L!}\Bigl[{\cal K}_l^{\cdot}{\cal H}^{\cdot}_{\nu_1}
\ldots{\cal H}^{\cdot}_{\nu_{L-1}}\Bigr]^{\{\mu_1\ldots\mu_L\}}
I_{N-1,N-l}^{\nu_1\ldots\nu_{L-1}}
\end{eqnarray}
To Eq.~(\ref{EQgeneraltred0}) 
\begin{eqnarray}
K_N^{\mu_1\ldots\mu_L}&=&{\cal W}^{\mu_1}\ldots 
{\cal W}^{\mu_L}I_N^n+
\sum\limits_{k=1}^L\Bigl[{\cal W}_{(L-k)}^{\cdot}{\cal K}_{l_1}^{\cdot}
\ldots{\cal K}_{l_k}^{\cdot}\Bigr]^{\{\mu_1\ldots\mu_L\}}I_{N-k,N-l_1,
\ldots,N-l_k}^n\nonumber
\end{eqnarray}
we apply Eq.~(\ref{Ik}) 
\begin{eqnarray}
I_{N-k,N-l_1,\ldots,N-l_k}^n
&=& \sum\limits_{j=0}^{k-1} (-1)^{k-j-1}\frac{2^{j}}{j!}
\Bigl[v_{\cdot}^{(k-j-1)}\,
r_{\cdot\nu_1}\ldots
r_{\cdot\nu_{j}}\Bigr]_{\{l_1\ldots l_{k-1}\}}
I_{N-1,N-l_k}^{\nu_1\ldots\nu_{j}} \nonumber
\label{IkAP}
\end{eqnarray}
and get, by taking into account
 combinatorial factors from contracting the tensor brackets
\begin{eqnarray}\label{EQrepAP}
K_N^{\mu_1\ldots\mu_L}&=&{\cal W}^{\mu_1}\ldots 
{\cal W}^{\mu_L}I_N^n
+\sum\limits_{k=1}^L\frac{1}{k}\sum\limits_{j=0}^{k-1}(-1)^{k-j-1}
\left(\begin{array}{l}L-j-1\\k-j-1\end{array}\right)\nonumber\\
&&\cdot\frac{2^j}{j!}
\Bigl[{\cal W}_{(L-j-1)}^{\cdot}{\cal K}_l^{\cdot}{\cal H}^{\cdot}_{\nu_1}\ldots
{\cal H}^{\cdot}_{\nu_j}\Bigr]^{\{\mu_1\ldots\mu_L\}}
I_{N-1,N-l}^{\nu_1\ldots\nu_j}
\end{eqnarray}
Using the fact that
\begin{eqnarray}
\sum\limits_{k=1}^{L}\sum\limits_{j=0}^{k-1} a_{kj}=
\sum\limits_{k=1}^{L}\sum\limits_{j=0}^{L-1} \theta(j\le k-1) a_{kj}=
a_{L,L-1} + 
\sum\limits_{j=0}^{L-2}\sum\limits_{k=j+1}^{L} a_{kj}
\end{eqnarray}
we can separate terms with and without ${\cal W}^\cdot$-vectors.
\begin{eqnarray}\label{EQsepAP}
K_N^{\mu_1\ldots\mu_L}&=&{\cal W}^{\mu_1}\ldots 
{\cal W}^{\mu_L}I_N^n
+\sum\limits_{k=1}^L\sum\limits_{j=0}^{L-2} 
\theta(j\le k-1)
\frac{2^j}{j!}\frac{(-1)^{k-j-1}}{k}
\left(\begin{array}{l}L-j-1\\k-j-1\end{array}\right)\\&&
\cdot\Bigl[{\cal W}_{(L-j-1)}^{\cdot}{\cal H}^{\cdot}_{\nu_1}\ldots
{\cal H}^{\cdot}_{\nu_j}{\cal K}_l^{\cdot}\Bigr]^{\{\mu_1\ldots\mu_L\}}
I_{N-1,N-l}^{\nu_1\ldots\nu_j}\nonumber \\
&&+\frac{2^{L-1}}{L!}
\Bigl[{\cal H}^{\cdot}_{\nu_1}\ldots
{\cal H}^{\cdot}_{\nu_{L-1}}{\cal K}_l^{\cdot}\Bigr]^{\{\mu_1\ldots\mu_L\}}
I_{N-1,N-l}^{\nu_1\ldots\nu_{L-1}}\nonumber
\end{eqnarray}
Now one has to rearrange the terms containing ${\cal W}$-vectors.\\
First we write $1/k$ in (\ref{EQsepAP}) as $1/k=(L-k)/(Lk) + 1/L$.
The $1/L$ part vanishes since the sum over $k$ just represents $(1-1)^{L-j-1}=0$. 
The $(L-k)/(Lk)$ part makes the upper bound of the $k$-sum 
to be $L-1$. From the tensor bracket a ${\cal W}^\cdot$
can be factored in a symmetric way:
\begin{eqnarray}
\Bigl[{\cal W}_{(L-j-1)}^{\cdot}{\cal H}^{\cdot}_{\nu_1}\ldots
{\cal H}^{\cdot}_{\nu_j}{\cal K}_l^{\cdot}\Bigr]^{\{\mu_1\ldots\mu_L\}}&=&
\frac{{\cal W}^{\mu_1}}{L-j-1}\Bigl[{\cal W}_{(L-j-2)}^{\cdot}{\cal H}^{\cdot}_{\nu_1}\ldots
{\cal H}^{\cdot}_{\nu_j}{\cal K}_l^{\cdot}\Bigr]^{\{\mu_2\ldots\mu_L\}}
\nonumber\\
&&+ \dots\nonumber\\
&\vdots&  \nonumber\\
&&+\frac{{\cal W}^{\mu_L}}{L-j-1}
\Bigl[{\cal W}_{(L-j-2)}^{\cdot}{\cal H}^{\cdot}_{\nu_1}\ldots
{\cal H}^{\cdot}_{\nu_j}{\cal
K}_l^{\cdot}\Bigr]^{\{\mu_1\ldots\mu_{L-1}\}}\nonumber
\end{eqnarray}
Hence Eq.~(\ref{EQsepAP}) is equivalent to
\begin{eqnarray}
K_N^{\mu_1\ldots\mu_L}-\frac{2^{L-1}}{L!}
\Bigl[{\cal H}^{\cdot}_{\nu_1}\ldots
{\cal H}^{\cdot}_{\nu_{L-1}}{\cal K}_l^{\cdot}\Bigr]^{\{\mu_1\ldots\mu_L\}}
I_{N-1,N-l}^{\nu_1\ldots\nu_{L-1}} \hspace{5cm}\nonumber\\
=\frac{{\cal W}^{\mu_1}}{L} 
\Bigl( {\cal W}^{\mu_2}\dots {\cal W}^{\mu_L} I_N^n +
\sum\limits_{k=1}^{L-1}\sum\limits_{j=0}^{k-1}
\frac{2^j}{j!}\frac{(-1)^{k-j-1}}{k}
\left(\begin{array}{l}L-j-2\\k-j-1\end{array}\right) \hspace{2cm}\nonumber\\
\cdot\Bigl[{\cal W}_{(L-j-2)}^{\cdot}{\cal H}^{\cdot}_{\nu_1}\ldots
{\cal H}^{\cdot}_{\nu_j}{\cal K}_l^{\cdot}\Bigr]^{\{\mu_2\ldots\mu_L\}}
I_{N-1,N-l}^{\nu_1\ldots\nu_j}\Bigr) + (L-1)\, \mbox{permutations} \label{this}
\end{eqnarray}
Comparing with Eq.~(\ref{EQrepAP}) we see that the right-hand side of Eq.~(\ref{this}) 
is just 
\begin{eqnarray}\label{EQcompAP}
&&\frac{{\cal W}^{\mu_1}}{L} K_N^{\mu_2\dots\mu_{L}} + (L-1)\, \mbox{permutations}
= \frac{1}{L}\Bigl[{\cal W}^{\cdot} 
K_N^{\{L-1 \,\mathrm{dots}\}}
\Bigr]^{\{\mu_1\ldots\mu_L\}}
\end{eqnarray}
Combining  (\ref{this}) and (\ref{EQcompAP}) we obtain  
the recursion relation (\ref{recAP}).

\section{Explicit reduction  for N=2,3,4,5,6}\label{appexplicit}

We now give explicit formulas for tensor integrals up to $N=6$, and rank $L\le N$. 
They have been derived by applying Eqs.~(\ref{recwithJ}), (\ref{EQshift}) and 
(\ref{36b}). 
The reduction stops because tadpole integrals of massless propagators
are zero, $I_1,I_1^{\mu}=0$. This shows that any tensor integral
with an arbitrary number of legs can be expressed in terms of scalar integrals
only. For non-exceptional kinematics
 higher dimensional integrals $I^{n+2m}_{N^{\prime}}\,(m=1,2)$ occur only with 
$N^{\prime}\le 4$ in the 
reduction of arbitrary $N$-point tensor integrals, as has been explained above.
All the higher dimensional integrals can be re-mapped to $n$-dimensional 
integrals with the scalar reduction formulas (\ref{EQSR2}), (\ref{EQSR3}) and 
(\ref{EQSR4}). Explicitly, the expressions for the higher dimensional 
integrals are given by

\begin{eqnarray}\label{EQhigherdimterms}
J^{\mu_1\mu_2}_{N=2,3,4} &=& (-1) \left( g^{\mu_1\mu_2}/2 - 
{\cal H}^{\mu_1\mu_2} \right) I_N^{n+2}\\
J^{\mu_1\mu_2\mu_3}_{N=3,4}\, &=& (-1) 
\Bigl[ \left( g/2 - {\cal H} \right)^{\cdot\cdot} 
{\cal W}^{\cdot}  \Bigr]^{\{\mu_1\mu_2\mu_3\}} I_N^{n+2} - 
\Bigl[ \left( g/2 - {\cal H} \right)^{\cdot\cdot} 
{\cal K}_l^{\cdot}\Bigr]^{\{\mu_1\mu_2\mu_3\}} I_{N-1,N-l}^{n+2}
\nonumber\\
J^{\mu_1\mu_2\mu_3\mu_4}_4 &=&  
\Bigl[ \left( g/2 - {\cal H} \right)_{(2)}^{\cdot\cdot} \Bigr]^{\{\mu_1\mu_2\mu_3\mu_4\}} 
I_4^{n+4}
- \Bigl[ \left( g/2 - {\cal H} \right)^{\cdot\cdot} 
{\cal  W}_{(2)}^{\cdot}\Bigr]^{\{\mu_1\mu_2\mu_3\mu_4\}} 
I_4^{n+2}\nonumber\\
&&- \frac{1}{2}\Bigl[ \left( g/2 - {\cal H} \right)^{\cdot\cdot} 
{\cal  W}^{\cdot}
{\cal  K}_l^{\cdot}\Bigr]^{\{\mu_1\mu_2\mu_3\mu_4\}} 
I_{3,4-l}^{n+2}
- \Bigl[ \left( g/2 - {\cal H} \right)^{\cdot\cdot} 
{\cal H}^{\cdot}_{\nu} {\cal  K}^{\cdot}_l\Bigr]^{\{\mu_1\mu_2\mu_3\mu_4\}} 
I_{3,4-l}^{\nu,n+2}\nonumber
\end{eqnarray} 
Note that rank one, $(n+2)$-dimensional three-point functions appear.
As the reduction rules do not depend on the space-time dimension
all the formulas given above
are also valid for higher dimensional tensor integrals.
The tensor structure is carried by the well-defined 4-dimensional objects 
${\cal W}^{\mu}, {\cal K}_l^{\mu}, {\cal H}^{\mu\nu}$ given in 
Eq.~(\ref{Lorentz}). Note that metric tensors occur only in the combination 
$( g/2 - {\cal H})$ 
as coefficients of the higher dimensional integrals $I_N^{n+2m}$. \\
For $N\ge 5$, the equal signs are only valid up 
to ${\cal O}(\epsilon)$ since the $J_N^{\mu_1\ldots \mu_L}$ terms have been 
dropped. \\ 
The shift  of the integrals 
$I_{N-1,N-l}^{\nu_1\ldots\nu_L}(R)$ 
has been done explicitly  up to $N=3$ \,in order to give examples for
the algorithm defined in (\ref{EQshift}). How to proceed for $N>3$ then should be 
obvious. \\
The reduction rules  can be implemented easily in
algebraic manipulation programs. 

\subsection*{N=2:}
Necessary for a non-vanishing result is that the external momentum
is not light-like. The Gram matrix is trivial. One has
${\cal W}^{\mu}(r)=r^\mu/2$, ${\cal H}^{\mu_1\mu_2}=r^{\mu_1}r^{\mu_2}/(2 r\cdot r)$.
\begin{eqnarray}
I^{\mu_1}_2(r)      &=& \frac{r^{\mu_1}}{2}\, I_2^n(r) \nonumber\\
I^{\mu_1\mu_2}_2(r) &=& \frac{1}{4(n-1)}\left(n\,r^{\mu_1}r^{\mu_2}-r^2 g^{\mu_1\mu_2}\right)\,I_2^n(r) 
\end{eqnarray}

\subsection*{N=3:}
We use the short-hand notation for the arguments of the
tensors and integrals defined in Eq.~(\ref{EQargumentvectors}) 
in the following: $R=[r_1,r_2,0]$, $R_{[1]}=[r_2-r_1,0]$, $\hat R_{[1]}=[r_2,0]$, 
etc. 
The arguments for the tensors ${\cal W}^{\mu}, {\cal K}_l^{\mu}, {\cal H}^{\mu\nu}$ 
are not written explicitly, e.g.
it is always understood that ${\cal W}^{\mu}={\cal W}^{\mu}(R)$, etc. 
\begin{eqnarray}
I^{\mu}_3(R) &=& {\cal W}^{\mu} \, I_3^n(R) 
+\sum\limits_{l=1}^{2} {\cal K}^{\mu}_l \left( I^n_2(R_{[l]})-I^n_2(\hat R_{[l]})  \right) \nonumber\\  
I^{\mu_1\mu_2}_3(R) 
&=&\frac{1}{2}\Bigl({\cal W}^{\mu_1} \, I_3^{\mu_2}(R)+{\cal W}^{\mu_2} \, I_3^{\mu_1}(R)\Bigr)\nonumber\\
&&+\sum\limits_{l=1}^{2}\left({\cal K}^{\mu_1}_l {\cal H}^{\mu_2}_{\nu}+{\cal K}^{\mu_2}_l {\cal H}^{\mu_1}_{\nu}\right)\left(r_l^{\nu}I_2^n(R_{[l]})+I_2^{\nu}(R_{[l]})-I_2^{\nu}(\hat R_{[l]})\right)\nonumber\\
&&-(g^{\mu_1\mu_2}/2-{\cal H}^{\mu_1\mu_2})I_3^{n+2}(R)\nonumber\\
&&\nonumber\\
I^{\mu_1\mu_2\mu_3}_3(R)  &=&\frac{1}{3}\Bigl(
{\cal W}^{\mu_1} \,\{I_3^{\mu_2\mu_3}-J_3^{\mu_2\mu_3}\}+
{\cal W}^{\mu_2} \,\{I_3^{\mu_1\mu_3}-J_3^{\mu_1\mu_3}\}+
{\cal W}^{\mu_3} \,\{I_3^{\mu_1\mu_2}-J_3^{\mu_1\mu_2}\}\Bigr)\nonumber\\
&&+\frac{2}{3}\sum\limits_{l=1}^{2}\Bigl[ {\cal K}^{\cdot}_l{\cal H}^{\cdot}_{\nu_1} {\cal H}^{\cdot}_{\nu_2}\Bigr]^{\{\mu_1\mu_2\mu_3\}}\Bigl(r_l^{\nu_1}r_l^{\nu_2}I_2^n(R_{[l]})+ r_l^{\nu_1}I_2^{\nu_2}(R_{[l]})+ r_l^{\nu_2}I_2^{\nu_1}(R_{[l]})\nonumber\\
&&+I_2^{\nu_1\nu_2}(R_{[l]})-I_2^{\nu_1\nu_2}(\hat R_{[l]})\Bigr)\nonumber\\
&&-\Bigl[ \left( g/2 - {\cal H} \right)^{\cdot\cdot} {\cal W}^{\cdot}  \Bigr]^{\{\mu_1\mu_2\mu_3\}} I_3^{n+2}(R) - 
\Bigl[ \left( g/2 - {\cal H} \right)^{\cdot\cdot} {\cal K}_l^{\cdot}\Bigr]^{\{\mu_1\mu_2\mu_3\}} I_{2,3-l}^{n+2}(R)
\end{eqnarray}

\subsection*{N=4:}

$R=[r_1,r_2,r_3,0]$. The integrals $I_{3,4-l}^{\nu_1\ldots\nu_L}$ are given in 
unshifted form. The integrals $J_4^{\mu_1\ldots\mu_L}$ are given in 
Eq.~(\ref{EQhigherdimterms}).
\begin{eqnarray}
I^{\mu_1}_4(R) &=& {\cal W}^{\mu_1} \, I_4^n(R) 
+\sum\limits_{l=1}^{3} {\cal K}^{\mu_1}_l I_{3,4-l}^n(R) \nonumber\\  
I^{\mu_1\mu_2}_4(R) &=& \frac{1}{2}\Bigl({\cal W}^{\mu_1} \, I_4^{\mu_2}(R)+{\cal W}^{\mu_2} \, I_4^{\mu_1}(R)\Bigr)+\sum\limits_{l=1}^{3}\Bigl[ {\cal K}^{\cdot}_l{\cal H}^{\cdot}_{\nu} \Bigr]^{\{\mu_1\mu_2\}}
I_{3,4-l}^{\nu}(R)+J_4^{\mu_1\mu_2}\nonumber \\
&&\nonumber\\
I^{\mu_1\mu_2\mu_3}_4(R) &=& \frac{1}{3}\Bigl(
{\cal W}^{\mu_1} \,\{I_4^{\mu_2\mu_3}-J_4^{\mu_2\mu_3}\}+
{\cal W}^{\mu_2} \,\{I_4^{\mu_1\mu_3}-J_4^{\mu_1\mu_3}\}+
{\cal W}^{\mu_3} \,\{I_4^{\mu_1\mu_2}-J_4^{\mu_1\mu_2}\}\Bigr)\nonumber\\
&&+ \frac{2}{3}\sum\limits_{l=1}^{3}\Bigl[ {\cal K}^{\cdot}_l{\cal H}^{\cdot}_{\nu_1}{\cal H}^{\cdot}_{\nu_2} \Bigr]^{\{\mu_1\mu_2\mu_3\}} I^{\nu_1\nu_2}_{3,4-l}(R) + J_4^{\mu_1\mu_2\mu_3}\nonumber \\
&&\nonumber\\
I^{\mu_1\mu_2\mu_3\mu_4}_4(R) &=&\frac{1}{4}\Bigl(
{\cal W}^{\mu_1} \,\{I_4^{\mu_2\mu_3\mu_4}-J_4^{\mu_2\mu_3\mu_4}\}+
{\cal W}^{\mu_2} \,\{I_4^{\mu_1\mu_3\mu_4}-J_4^{\mu_1\mu_3\mu_4}\}\nonumber\\
&&+{\cal W}^{\mu_3} \,\{I_4^{\mu_1\mu_2\mu_4}-J_4^{\mu_1\mu_2\mu_4}\}
+{\cal W}^{\mu_4} \,\{I_4^{\mu_1\mu_2\mu_3}-J_4^{\mu_1\mu_2\mu_3}\}\Bigr)\nonumber\\
&&+ \frac{1}{3}\sum\limits_{l=1}^{3}\Bigl[{\cal K}_l^{\cdot}{\cal H}^{\cdot}_{\nu_1}{\cal H}^{\cdot}_{\nu_2}{\cal H}^{\cdot}_{\nu_3}\Bigr]^{\{\mu_1\mu_2\mu_3\mu_4\}}I_{3,4-l}^{\nu_1\nu_2\nu_3}(R) + J_4^{\mu_1\mu_2\mu_3\mu_4} 
\end{eqnarray}


 \subsection*{N=5:}
As discussed above, no higher dimensional integrals appear 
in the reduction of tensor $N$-point integrals for $N>4$, as long as the
external momenta are non-exceptional.  They occur only implicitly
by reduction of 2,3,4-point tensor integrals. Now, $R=[r_1,r_2,r_3,r_4,0]$.
The formulas are valid up to 
${\cal O(\epsilon})$ since $J_5^{\{\mu\}}$ terms have been dropped.
\begin{eqnarray}
I^{\mu_1}_5(R) &=& {\cal W}^{\mu_1} \, I_5^n(R) 
+\sum\limits_{l=1}^{4} {\cal K}^{\mu_1}_l \,
 I^n_{4,5-l}(R) \nonumber\\  
I^{\mu_1\mu_2}_5(R) &=& \frac{1}{2}\Bigl({\cal W}^{\mu_1} \, I_5^{\mu_2}(R)+{\cal W}^{\mu_2} \, I_5^{\mu_1}(R)\Bigr)+\sum\limits_{l=1}^{4}\Bigl[{\cal K}^{\cdot}_l {\cal H}^{\cdot}_{\nu}\Bigr]^{\{\mu_1\mu_2\}}I_{4,5-l}^{\nu}(R) \nonumber\\
&&\nonumber\\
I^{\mu_1\mu_2\mu_3}_5(R) &=& \frac{1}{3}\Bigl(
{\cal W}^{\mu_1} \,I_5^{\mu_2\mu_3}+
{\cal W}^{\mu_2} \,I_5^{\mu_1\mu_3}+
{\cal W}^{\mu_3} \,I_5^{\mu_1\mu_2}\Bigr)\nonumber\\
&&+\frac{2}{3}\sum\limits_{l=1}^{4}\Bigl[ {\cal K}^{\cdot}_l{\cal H}^{\cdot}_{\nu_1} {\cal H}^{\cdot}_{\nu_2}\Bigr]^{\{\mu_1\mu_2\mu_3\}}I_{4,5-l}^{\nu_1\nu_2}(R)\nonumber\\
&&\nonumber\\
I^{\mu_1\mu_2\mu_3\mu_4}_5(R) &=& \frac{1}{4}\Bigl(
{\cal W}^{\mu_1} \,I_5^{\mu_2\mu_3\mu_4}+
{\cal W}^{\mu_2} \,I_5^{\mu_1\mu_3\mu_4}+{\cal W}^{\mu_3} \,I_5^{\mu_1\mu_2\mu_4}
+{\cal W}^{\mu_4} \,I_5^{\mu_1\mu_2\mu_3}\Bigr)\nonumber\\
&&+ \frac{1}{3}\sum\limits_{l=1}^{4}\Bigl[{\cal K}_l^{\cdot}{\cal H}^{\cdot}_{\nu_1}{\cal H}^{\cdot}_{\nu_2}{\cal H}^{\cdot}_{\nu_3}\Bigr]^{\{\mu_1\mu_2\mu_3\mu_4\}}I_{4,5-l}^{\nu_1\nu_2\nu_3}(R)\nonumber\\
&&\nonumber\\
I^{\mu_1\mu_2\mu_3\mu_4\mu_5}_5(R) &=& \frac{1}{5}\Bigl(
{\cal W}^{\mu_1} \,I_5^{\mu_2\mu_3\mu_4\mu_5}+
\ldots
+{\cal W}^{\mu_5} \,I_5^{\mu_1\mu_2\mu_3\mu_4}\Bigr)\nonumber\\
&&+ \frac{2}{15}\sum\limits_{l=1}^{4}\Bigl[{\cal K}_l^{\cdot}{\cal H}^{\cdot}_{\nu_1}{\cal H}^{\cdot}_{\nu_2}{\cal H}^{\cdot}_{\nu_3}{\cal H}^{\cdot}_{\nu_4}\Bigr]^{\{\mu_1\mu_2\mu_3\mu_4\mu_5\}}I_{4,5-l}^{\nu_1\nu_2\nu_3\nu_4}(R) 
\end{eqnarray}

\subsection*{N=6:}
For non-exceptional kinematics any set of four vectors out of 
$\{r_1,r_2,r_3,r_4,r_5\}$ spans 
4-dimensional Minkowski space. We will express the tensor 6-point functions
in terms of scalar 6-point functions and pentagon integrals. 
The latter then can be  reduced further by using (\ref{EQSR6}) and the reduction 
formulas for $N=5$ with the corresponding new argument vectors. 
\begin{eqnarray}
I^{\mu_1}_6(R) &=& {\cal W}^{\mu_1} \, I_6^n(R) 
+\sum\limits_{l=1}^{5} {\cal K}^{\mu_1}_l\,  I^n_{5,6-l}(R) \nonumber\\  
I^{\mu_1\mu_2}_6(R) &=&\frac{1}{2}\Bigl({\cal W}^{\mu_1} \, I_6^{\mu_2}(R)+{\cal W}^{\mu_2} \, I_6^{\mu_1}(R)\Bigr)+\sum\limits_{l=1}^{5}\Bigl[{\cal K}^{\cdot}_l {\cal H}^{\cdot}_{\nu}\Bigr]^{\{\mu_1\mu_2\}}I_{5,6-l}^{\nu}(R)\nonumber\\
&&\nonumber\\
I^{\mu_1\mu_2\mu_3}_6(R) &=& \frac{1}{3}\Bigl(
{\cal W}^{\mu_1} \,I_6^{\mu_2\mu_3}+
{\cal W}^{\mu_2} \,I_6^{\mu_1\mu_3}+
{\cal W}^{\mu_3} \,I_6^{\mu_1\mu_2}\Bigr)\nonumber\\
&&+\frac{2}{3}\sum\limits_{l=1}^{5}\Bigl[ {\cal K}^{\cdot}_l{\cal H}^{\cdot}_{\nu_1} {\cal H}^{\cdot}_{\nu_2}\Bigr]^{\{\mu_1\mu_2\mu_3\}}I_{5,6-l}^{\nu_1\nu_2}(R)\nonumber\\
&&\nonumber\\
I^{\mu_1\mu_2\mu_3\mu_4}_6(R) &=& \frac{1}{4}\Bigl(
{\cal W}^{\mu_1} \,I_6^{\mu_2\mu_3\mu_4}+
{\cal W}^{\mu_2} \,I_6^{\mu_1\mu_3\mu_4}+{\cal W}^{\mu_3} 
\,I_6^{\mu_1\mu_2\mu_4}
+{\cal W}^{\mu_4} \,I_6^{\mu_1\mu_2\mu_3}\Bigr)\nonumber\\
&&+ \frac{1}{3}\sum\limits_{l=1}^{5}\Bigl[{\cal K}_l^{\cdot}
{\cal H}^{\cdot}_{\nu_1}{\cal H}^{\cdot}_{\nu_2}{\cal H}^{\cdot}_{\nu_3}
\Bigr]^{\{\mu_1\mu_2\mu_3\mu_4\}}I_{5,6-l}^{\nu_1\nu_2\nu_3}(R)\nonumber\\
&&\nonumber\\
I^{\mu_1\mu_2\mu_3\mu_4\mu_5}_6(R) &=& \frac{1}{5}\Bigl(
{\cal W}^{\mu_1} \,I_6^{\mu_2\mu_3\mu_4\mu_5}+
\ldots
+{\cal W}^{\mu_5} \,I_6^{\mu_1\mu_2\mu_3\mu_4}\Bigr)\nonumber\\
&&+ \frac{2}{15}\sum\limits_{l=1}^{5}\Bigl[{\cal K}_l^{\cdot}
{\cal H}^{\cdot}_{\nu_1}{\cal H}^{\cdot}_{\nu_2}{\cal H}^{\cdot}_{\nu_3}
{\cal H}^{\cdot}_{\nu_4}\Bigr]^{\{\mu_1\mu_2\mu_3\mu_4\mu_5\}}
I_{5,6-l}^{\nu_1\nu_2\nu_3\nu_4}(R)\nonumber\\
&&\nonumber\\
I^{\mu_1\mu_2\mu_3\mu_4\mu_5\mu_6}_6(R) &=& \frac{1}{6}\Bigl(
{\cal W}^{\mu_1} \,I_6^{\mu_2\mu_3\mu_4\mu_5\mu_6}+
\ldots+{\cal W}^{\mu_6} \,I_6^{\mu_1\mu_2\mu_3\mu_4\mu_5}\Bigr)\nonumber\\
&&+ \frac{2}{45}\sum\limits_{l=1}^{5}
\Bigl[{\cal K}_l^{\cdot}{\cal H}^{\cdot}_{\nu_1}{\cal H}^{\cdot}_{\nu_2}
{\cal H}^{\cdot}_{\nu_3}{\cal H}^{\cdot}_{\nu_4}{\cal H}^{\cdot}_{\nu_5}
\Bigr]^{\{\mu_1\mu_2\mu_3\mu_4\mu_5\mu_6\}}
I_{5,6-l}^{\nu_1\nu_2\nu_3\nu_4\nu_5}(R)\nonumber
\end{eqnarray}


\begin{thebibliography}{99}
\bibitem{treelevelamplitudes} M.~Mangano, S.~Parke, Phys. Rep. 200 (1991) 301;\\
P.~Draggiotis, R.H.P.~Kleiss, C.G.~Papadopoulos, Phys. Lett. B439 (1998) 157;\\
F.~Caravaglios, M.~Mangano, M.~Moretti, Nucl. Phys. B539, (1999) 215.
        
\bibitem{EllisSexton} R.K.~Ellis and J.~Sexton, Nucl. Phys. B269, 445 (1986).
\bibitem{BernDixonKosower} Z.~Bern, L.~Dixon, D.A.~Kosower, 
Phys. Lett. B302 (1993) 299;  Nucl. Phys. B412 (1994) 751.
\bibitem{KunsztTrocsanyi} Z. Kunszt, A. Signer, Z. Tr\'ocs\'anyi, 
                          Phys. Lett. B336 (1994) 529.
\bibitem{WeinzierlKosower} S.~Weinzierl, D.A.~Kosower, hep/ph-9901277.
\bibitem{CampbellCullenGlover} J.~M.~Campbell, M.~A.~Cullen, E.W.N.~Glover, 
          Eur. Phys. J.C9 (1999) 245.
\bibitem{PassarinoVeltman} G.~Passarino, M.~Veltman, Nucl. Phys. B160 (1979), 151.
\bibitem{Oldenbourg} G.J.~van Oldenbourg, J.A.M. Vermaseren, 
                     Z. Phys. C46 (1990) 425.
\bibitem{VanNeervenVermaseren} W.L.~van~Neerven, J.A.M. Vermaseren, 
                               Phys. Lett. B137 (1984) 241.
\bibitem{Signer} A.~Signer, Ph.D. thesis, Diss. ETH Nr. 11143, Z\"urich 1995.
\bibitem{Davydychev} A.I.~Davydychev, Phys. Lett. B263 (1991) 107.
\bibitem{FleischerJegerlehnerTarasov} 
O.V.~Tarasov, Phys. Rev. D54 (1996) 6479;\\
J.~Fleischer, F.~Jegerlehner, O.V.~Tarasov hep-ph/9907327;
\bibitem{DevarajStuart} G.~Devaraj, R.G.~Stuart, Nucl. Phys. B519, (1998) 483;\\
                        R.G.~Stuart, Comput. Phys. Commun. 48 (1988) 367.  
\bibitem{Pittau} R.~Pittau, Comput. Phys. Commun. 104 (1997) 23; 
                            ibid. 111 (1998) 48.
\bibitem{Weinzierl} S.~Weinzierl, Phys. Lett. B450, (1999) 234.
\bibitem{Denner} A.~Denner, Fortsch.Phys. 41, (1993) 307.
\bibitem{CampbellGloverMiller} J.~M.~Campbell, E.W.N.~Glover, D.J.~Miller, 
                               Nucl. Phys. B498 (1997) 397.
\bibitem{Melrose} D.B.~Melrose, Il Nouvo Cimento 40A (1965) 181.  
\bibitem{Koecher} M.~K\"ocher, Lineare Algebra und analytische Geometrie 
(2nd edition), \\Springer--Verlag, Berlin (1985); \\
 A.~Ben-Israel, Th.N.E.~Greville, Generalized Inverses, J.~Wiley, 
 New York (1974).
\bibitem{tHooftVeltman} G.'t~Hooft, M.~Veltman, Nucl. Phys. B153 (1979) 365.
\end{thebibliography}
\end{document}